\documentclass{article}
\usepackage[utf8]{inputenc}
\usepackage{amsmath}
\usepackage{amsthm}
\usepackage{amssymb}
\usepackage{bbm}
\usepackage[round]{natbib}
\usepackage[english]{babel} 
\usepackage{enumerate}
\usepackage[margin=1in]{geometry}
\usepackage{graphicx}
\usepackage{pdfpages}
\usepackage{makecell}
\usepackage{cases}
\usepackage{caption}
\usepackage{subcaption}
\usepackage[normalem]{ulem}
\usepackage{xr-hyper} 
\usepackage{hyperref}

\usepackage{float}
\floatstyle{plaintop}
\restylefloat{table}

\hypersetup{
    colorlinks=false, 
    linktoc=all,     
    linkcolor=blue,  
    hidelinks,
}

\DeclareMathOperator{\rank}{rank}
\DeclareMathOperator{\diag}{diag}
\DeclareMathOperator{\Var}{Var}
\DeclareMathOperator{\Cov}{Cov}

\DeclareMathOperator{\Exp}{E}
\DeclareMathOperator{\logit}{logit}
\DeclareMathOperator{\probit}{probit}
\DeclareMathOperator{\cauchit}{cauchit}
\DeclareMathOperator{\cloglog}{cloglog}

\newtheorem*{theorem*}{Theorem} 

\newtheorem{lemma}{Lemma}

\newcommand{\reals}{\mathbb{R}}

\newcommand\Tstrut{\rule{0pt}{2.6ex}}         

\makeatletter
\newcommand*{\addFileDependency}[1]{
  \typeout{(#1)}
  \@addtofilelist{#1}
  \IfFileExists{#1}{}{\typeout{No file #1.}}
}
\makeatother

\newcommand*{\myexternaldocument}[1]{%
    \externaldocument{#1}%
    \addFileDependency{#1.tex}%
    \addFileDependency{#1.aux}%
}

\myexternaldocument{supplementary_main}

\begin{document}
\title{\vspace{-10mm} \textbf{\Large{A Generalized Hosmer-Lemeshow Goodness-of-Fit Test for a Family of Generalized Linear Models}}}
\author{\textbf{\normalsize{Nikola Surjanovic, Richard Lockhart, and Thomas M. Loughin}}}
\date{\normalsize{Department of Statistics and Actuarial Science, Simon Fraser University, Burnaby, British Columbia V5A 1S6, Canada}}
\maketitle
\baselineskip=14pt 

\section*{\large{Summary}}
{\small Generalized linear models (GLMs) are used within a vast number of application domains. However, formal goodness of fit (GOF) tests for the overall fit of the model---so-called ``global'' tests---seem to be in wide use only for certain classes of GLMs. In this paper we develop and apply a new global goodness-of-fit test, similar to the well-known and commonly used Hosmer-Lemeshow (HL) test, that can be used with a wide variety of GLMs. The test statistic is a variant of the HL test statistic, but we rigorously derive an asymptotically correct sampling distribution of the test statistic using methods of \cite{stute2002model}. Our new test is relatively straightforward to implement and interpret. We demonstrate the test on a real data set, and compare the performance of our new test with other global GOF tests for GLMs, finding that our test provides competitive or comparable power in various simulation settings. Our test also avoids the use of kernel-based estimators, used in various GOF tests for regression, thereby avoiding the issues of bandwidth selection and the curse of dimensionality. Since the asymptotic sampling distribution is known, a bootstrap procedure for the calculation of a p-value is also not necessary, and we therefore find that performing our test is computationally efficient.

\noindent \textit{Key Words: } empirical regression process, exponential dispersion family, generalized linear model, goodness-of-fit test, Hosmer-Lemeshow test}

\section{Introduction}

Generalized linear models (GLMs), which include the linear, logistic, and Poisson regression models, among others, have been used within a vast number of application domains, and are extremely popular in medical and biological applications. For an application and some examples illustrating the use of GLMs, see \cite{lemeshow1988predicting} and \cite{agresti2015foundations}, for example. One of the most common GLMs, the logistic regression model, is often used when a binary response is to be regressed upon one or more explanatory variables. For example, this may occur when the response represents the survival or death of a patient, and the explanatory variables might be characteristics of the individual or various treatment methods. Other GLMs, such as the Poisson regression model, can be used to model count data. This can occur, for example, when the response is a count of different species of hummingbirds captured during mist-netting sessions in a cloud forest \citep{becker2020hummingbird}.

Naturally, it is desirable to have a model that fits the observed data well. Insight as to whether a model is a poor fit for a given dataset can be provided by a variety of measures, including goodness-of-fit (GOF) tests, as described by \cite{bilder2014analysis}. In particular, global (omnibus) goodness-of-fit tests inspect the validity of the GLM model as a whole, with the drawback of not providing much insight into the source of any potential deviation from the null model.

\cite{hosmer1980goodness} constructed a global GOF test for the logistic regression model, the Hosmer-Lemeshow (HL) test, which applies a Pearson test statistic to differences of observed and expected event counts from data grouped based on the ordered fitted values from the model. As a result, their test is very easy to interpret and is extremely popular, particularly in medical applications. Due to the simplicity of the test, it is tempting to apply it to other GLMs. However, as simulation results in Section \ref{sec:sim_results_application} suggest, direct application of the HL test to GLMs with many parameters and a non-binomial response results in a distribution of the test statistic that is not adequately represented by a chi-squared distribution in finite samples, thereby adversely affecting the type 1 error rate and power of the test.

A considerable number of the global GOF tests available for GLMs with a non-binomial or non-normal response (or other regression models beyond GLMs) involve kernel density estimation or some other form of smoothing---for example, the test of \cite{cheng1994testing}. However, \cite{gonzalez2013updated} mention that the selection of the smoothing parameter used in some tests is ``a broadly studied problem in regression estimation but with serious gaps for testing problems''. Also, the introduction of continuous covariates in a GLM model can render certain basic tests, such as the Pearson chi-squared test, invalid, due to issues that arise \citep[see][]{pulkstenis2002two}. In this paper, we address these issues and develop an omnibus GOF test for many GLMs that is both easy to compute and adaptable to many types of covariates. Specifically, we derive a modification to the HL test statistic that allows it to be generalized to a wide variety of GLMs. The modification is based on application of theory developed by \cite{stute2002model}, who work in a very general framework. We show that the test statistic has an asymptotic chi-squared distribution for many GLMs in the exponential dispersion family, adding to the appealing simplicity of the test.   

We begin Section \ref{sec:bg_notation} with an overview of previously developed global GOF tests. Section \ref{sec:methods} provides a more detailed description of certain existing global GOF tests, along with our new test. The design for our simulation study comparing these tests is laid out in Section \ref{sec:sim_design}, and the results are provided in Section \ref{sec:sim_results_application}. An application to a dataset is also performed in Section \ref{sec:sim_results_application}. We find that our test provides competitive or comparable power in various simulation settings, is computationally efficient, and avoids the use of kernel-based estimators. Finally, we discuss the results and potential future work in Section \ref{sec:discussion}. The proofs of several results are provided in the supplementary material.


\section{Background and Notation}
\label{sec:bg_notation}
Throughout this paper, we let $Y$ be a response variable that is associated with a covariate vector, $X$, where $X \in \mathbb{R}^d$. We write $(X_i, Y_i)$, $i=1,\ldots,n$, to denote a random sample where each $(X_i, Y_i)$ has the same distribution as $(X,Y)$, and provides observed data $(x_i,y_i)$. 

The Hosmer-Lemeshow GOF test assesses departures between observed and expected event counts from data grouped based on the fitted values from the model. For the moment, we restrict our attention to binary response logistic regression models, in which each $y_i \in \{0,1\}$, although the HL test can be used in a binomial setting by treating each $y_i$ as the count of successes resulting from all trials observed at the corresponding $x_i$. In the binary case, we assume that
\begin{equation*}
    \Exp(Y|X=x) = \pi(\beta^\top x) = \frac{\exp(\beta^\top x)}{1+\exp(\beta^\top x)},
\end{equation*}
for some $\beta \in \mathbb{R}^d$. The likelihood function is then given by
\begin{equation*}
    \mathcal{L}(\beta) = \prod_{i=1}^n \pi(\beta^\top x_i)^{y_i} (1-\pi(\beta^\top x_i))^{1-y_i},
\end{equation*}
from which a maximum likelihood estimate (MLE), $\beta_n$, of $\beta$ can be obtained. 

Computing the HL test statistic starts with partitioning the data into G groups. This is often done in a way so that the groups are of approximately equal size and fitted values within each group are similar. The partition fixes interval endpoints, $-\infty = k_0 < k_1 < \cdots < k_{G-1} < k_G = \infty$; the $k_g$ are often set to be equal to the logit of equally-spaced quantiles of the fitted values, $\hat\pi_i = \pi(\beta_n^\top x_i)$. Then, we define $I_i^{(g)} = \mathbbm{1}(k_{g-1} < \beta_n^\top x_i \leq k_g)$, $O_g = \sum_{i=1}^n y_i I_i^{(g)}$, $E_g = \sum_{i=1}^n \hat\pi_i I_i^{(g)}$, $n_g = \sum_{i=1}^n I_i^{(g)}$, and $\bar{\pi}_g = E_g/n_g$, where $\mathbbm{1}(A)$ is the indicator function on a set $A$. Then, $n_g$ represents the number of observations in the $g$th group, and $\bar{\pi}_g$ represents the average of the fitted values within the $g$th group. Finally, we can define the HL test statistic as
\begin{equation}
    \widehat{C}_G = \sum_{g=1}^G \frac{(O_g - E_g)^2}{n_g \bar{\pi}_g (1-\bar{\pi}_g)}.
\end{equation}

The theory behind the regular HL test, based on work by \cite{moore1975unified}, suggests that the asymptotic null distribution of the test statistic follows the distribution of a weighted sum of chi-squared random variables. \cite{hosmer1980goodness} approximated this complex distribution with a single chi-squared, where the degrees of freedom were determined partly by simulation. That this is indeed only an approximation of the \textit{asymptotic} null distribution can be clearly seen first in \cite{hosmer1980goodness}, \cite{lemeshow1982review}, and in many of the aforementioned papers. We show in our own simulations that a direct application of similarly constructed tests to other GLMs, such as the Poisson regression model, can perform poorly due to improper theoretical justification, and that our generalization of the HL test addresses this issue.

Another grouped GOF test for logistic regression models was introduced by \cite{tsiatis1980note} around the same time as the original HL test. This test partitions the covariate space, thereby grouping the data, from which a test statistic can be calculated. Expanding the definition of $I_i^{(g)}$ to more generically indicate that the $i$th observation is in the $g$th group, then
\begin{equation}
\label{eq:group_defns}
    I_i^{(g)} = 
    \begin{cases}
        \mathbbm{1}(k_{g-1} < \beta_n^\top x_i \leq k_g), & \text{HL test ($k_g$ random, chosen by the data)} \\
        \mathbbm{1}(x_i \in R_g), & \text{Tsiatis test ($R_g$ fixed and disjoint, $\cup_g R_g = \mathbb{R}^d$)}
    \end{cases}.
\end{equation}
The matrix of the quadratic form for the HL test statistic is diagonal, as seen later in Section \ref{sec:methods}, whereas for the Tsiatis test statistic it is not. Thus, if groups are formed in the same way for both tests, the main difference between the HL test and the Tsiatis test is that the latter accounts for correlations between groups of residuals. 

The Tsiatis test statistic, with covariate space partitionings defined before observing the data, can be easily transferred to other GLMs with a canonical link, such as the Poisson regression model with a log link. \cite{canary2016summary} constructed a generalized Tsiatis test statistic for binary regression models with a noncanonical link, and then made use of an HL-like partitioning scheme, as in the first case of (\ref{eq:group_defns}). Their approach is very interesting, but the theoretical validity of such a partitioning method for the generalized Tsiatis test should be verified, similar to what was done by \cite{halteman1980goodness} in the case of logistic regression. The generalized Tsiatis test statistic for binary regression models is equivalent to our proposed test statistic when both tests use the same partitioning scheme, i.e., either of the cases in (\ref{eq:group_defns}). We formalize the theoretical validity of the asymptotic chi-squared distribution for the generalized Tsiatis test statistic when the first grouping method of (\ref{eq:group_defns}) is used. We also allow for extensions to a much wider class of GLMs, as was briefly suggested by \cite{canary2013grouped}. 

Previous attempts have been made to generalize the HL test to models other than the logistic regression GLM. For example, \cite{blizzard2006parameter} and \cite{quinn2015goodness} apply the HL test to binomial regression models with a log link, and \cite{canary2016summary} consider several GOF tests, including the HL test, for binomial regression models with various noncanonical links. A version of the HL test has also been transferred to the multinomial regression model, discussed by \cite{fagerland2008multinomial}, and the proportional odds and other ordinal logistic regression models, as described by \cite{fagerland2013goodness, fagerland2016tests}.

Tests of fit have also been constructed that can be used with broader classes of GLM models. An early contribution came from \cite{su1991lack}, who considered the supremum of a multivariate process based on a cumulative sum of residuals. \cite{stute2002model} presented a test statistic based on the Cramér-von Mises statistic applied to a transformed residual process. Other GOF tests that can be used with GLMs can be found in \cite{cheng1994testing, lin2002model, liu2005testing, rodriguez1998testing, xiang1995testing}. A nice review of GOF tests for regression models is given by \cite{gonzalez2013updated}. While many of these tests appear to have merit, they do not seem to have been widely adopted in practice. Unfortunately, the p-values accompanying some tests need to be obtained through simulation. \cite{christensen2015lack} noted that with several explanatory variables, simulations required by the Su-Wei GOF test can be ``prohibitively time consuming''. Aside from computational efficiency, the use of smoothing-based methods in some tests raises the question of how these tests perform in higher-dimensional situations. Other concerns with some tests are the interpretability and ease of use, since some tests appear in mathematical context and do not have readily available software. As a consequence, such tests might be inaccessible to users in certain fields. We believe that our new test is fairly accessible, and we show that it works in high-dimensional settings and is computationally efficient.

To generalize the HL test to other GLMs, we use notation similar to that used by \cite{stute2002model}. We assume under the null hypothesis, given by (\ref{eq:H0glm}) below, that $\Exp(Y^2) < \infty$. As a consequence, $\Exp(Y)<\infty$, and we may define $m^*(x) = \Exp(Y |X = x)$, and $\sigma^2(x) =\Var(Y|X=x)$. We also assume that $\sigma^2(x)>0$ for all $x$.  

The GLM that we fit and test assumes that the conditional density of $Y$ given $X=x$ is an exponential family member with inverse link function, $m$. Specifically, the conditional density of $Y$ given $X=x$ has the form 
\begin{equation}
\label{eq:f_y|x}
    f_{Y|X}(y|x,\beta_0) = \exp\{y\theta - b(\theta)\} \nu(dy),
\end{equation}
where the vector $\beta_0$, $\theta$, and $x$ are related by $\Exp(Y|X=x)=b'(\theta)=m(\beta_0^T x)$, and $\nu$ is a suitable dominating measure. In such a model the variance is a function of the mean and we may write $\sigma^2(x) = v(m(\beta^\top x))$, for a smooth function $v$ determined by the function $b$. In some cases it is of interest to add an (unknown) dispersion parameter, $\phi$, to the model and assume further that there is a scalar $\phi_0$ such that the density of $Y$ given $X=x$ has the form
\begin{equation}
\label{eq:f_y|x_phi}
    f_{Y|X}(y|x,\beta_0,\phi_0) = \exp\left\{\frac{y\theta - b(\theta)}{\phi} - c(y,\phi)\right\}
\end{equation}
with respect to a (possibly different) dominating measure $\nu$, where $\beta_0$, $\phi_0$, $\theta$, and $x$ are related by similar equalities as before. For simplicity, we focus on the form given by (\ref{eq:f_y|x}), i.e., the dispersion parameter is known. We later offer a discussion on how to apply our GOF test to models with an unknown dispersion parameter.

We test the null hypothesis that
\begin{align}
\begin{split}
\label{eq:H0glm}
    m^*(x) &= m(\beta^\top x) \text{ for some $\beta \in \mathbb{R}^d$, and} \\
    Y|X &\sim f_{Y|X},
\end{split}
\end{align}
for all values of $x$ that are of interest, for a specified inverse link function $m(\cdot)$, where $\|\cdot\|$ denotes the Euclidean norm. We denote the maximum likelihood estimate of $\beta$ by $\beta_n$.

We specify in Section \ref{sec:validGLMs} conditions on $m(\cdot)$ and $f_{Y|X}$, and lay out more precisely to which GLMs our new GOF test may be applied. Also, note that our test is truly a global or ``omnibus'' test, since rejecting the null hypothesis (\ref{eq:H0glm}) could suggest that an incorrect link was specified, a variable is missing from the model, or that an entirely different model should be used. Such a test is in line with the global HL test, which does not test against any specific alternative.


\section{Methods and Test Statistics}
\label{sec:methods}
Tests for GOF require statistics that measure departures from the null hypothesis. To this end, define the true error, a random quantity, as $ \epsilon = Y - m(\beta_0^\top X) $. Also relevant to our test is the following process, for $u \in \mathbb{R}$, as defined in \cite{stute2002model}:
\begin{equation}
    R_n^1(u) = \frac{1}{\sqrt{n}} \sum_{i=1}^n \mathbbm{1}(\beta_n^\top X_i \leq u) [Y_i - m(\beta_n^\top X_i)]. 
\end{equation}

The HL test can be rewritten as a quadratic form, in terms of the residual process, $R_n^1(u)$. We first define the $G$-vector
\begin{equation}
\label{eq:S_n^1}
    S_n^1 = (R_n^1(k_1) - R_n^1(k_0), \ldots, R_n^1(k_G) - R_n^1(k_{G-1}) )^\top. 
\end{equation}
Then, the HL test statistic can be rewritten as $\widehat{C}_G = S_n^{1 \top} D^{-1} S_n^1$, where
\begin{equation}
\label{eq:D}
    D = \text{diag}\left( \frac{n_g \bar{\pi}_g(1-\bar{\pi}_g)}{n}\right).
\end{equation}

Provided that $G>d$---a requirement not always cited in references to the HL test---and upon verifying certain conditions of Theorem 5.1 in \cite{moore1975unified}, \cite{hosmer1980goodness} conclude that their test statistic is asymptotically distributed under the null hypothesis as a weighted sum of chi-squared random variables. That is,
$$ \widehat{C}_G \xrightarrow{d} \chi^2_{G-d} + \sum_{j=1}^d \lambda_j \chi_{1j}^2,$$
where each $\chi_{1j}^2$ is a chi-squared random variable with 1 degree of freedom, and each $\lambda_j$ is an eigenvalue of a particular matrix that depends on $\beta_0$ and the distribution of $X$. Then, through simulations, they conclude that the term $\sum_{j=1}^d \lambda_d \chi^2_{1j}$ can be approximated in various settings by a $\chi^2_{d-2}$ distribution. The results from the simulation study then led to the recommended $G-2$ degrees of freedom. In other words, $ \widehat{C}_G \stackrel{.}{\sim} \chi^2_{G-2}.$ However, in certain settings with a finite sample size this does not serve as a good approximation, as we discuss in Section \ref{sec:sim_results_application}.

\subsection{A Naive Generalization of the Hosmer-Lemeshow Test}
The HL test statistic depends on the binomial assumption only through $D$ in (\ref{eq:D}), with the $g$th diagonal element representing an estimate of the variance of the counts in the $g$th group, divided by $n$. To extend this test to other GLMs, it is tempting to define a ``naive'' HL test statistic
\begin{equation}
    \widehat{C}^*_G = S_n^{1 \top} (D^*)^{-1} S_n^1,
\end{equation}
where
$$ D^* = \diag\left( \frac{1}{n} \sum_{i=1}^n \widehat{\Var}(Y|X=x_i) \mathbbm{1}(k_{g-1} < \beta_n^\top x_i \leq k_g) \right),$$
similar to the estimates of the variances of group counts given in the original HL test through $D$. For example, for Poisson regression models one would use
$$ D^* = \diag\left( \frac{1}{n} \sum_{i=1}^n m(\beta_n^\top x_i) \mathbbm{1}(k_{g-1} < \beta_n^\top x_i \leq k_g) \right),$$
since the conditional variance of the response is equal to the conditional mean. This idea is very briefly suggested in \cite{agresti1996introduction} on p.\ 90, and in \cite{bilder2014analysis}. Implementing this test for Poisson regression models, our simulation results suggest that as the number of estimated parameters in the model increases, the mean and variance of the $\widehat{C}^*_G$ test statistic tend to decrease for a fixed sample size. Further properties of this test are discussed in Section \ref{sec:sim_results}.

\subsection{The Generalized HL Test Statistic}
\label{sec:genHL}
We now provide the test statistic for our generalized HL test. First, we define some important matrices. Let 
\begin{align*}
    \left(G_n^{*}\right)_{gi} &= \mathbbm{1}(k_{g-1} < \beta_n^\top x_i \leq k_g), \\
    V^{*1/2} &= \diag\left( \sigma(x_i) \right) \big\rvert_{\beta = \beta_0}, \\
    W^{1/2} &= \diag\left( m'(\beta^\top x_i) \cdot \frac{1}{\sigma(x_i)} \right)\Biggr\rvert_{\beta=\beta_0},
\end{align*}
for $i = 1, \ldots, n$, and $g = 1, \ldots, G$. Also, let $X^*$ be the $n \times d$ matrix whose $i$th row is given by $x_i^\top$, for $i=1,\ldots,n$. Define $W_n^{1/2}$ and $V_n^{*1/2}$ the same as $W^{1/2}$ and $V^{*1/2}$, respectively, but evaluated at $\beta_n$ instead of $\beta_0$. Note that if there is a dispersion parameter in the model then $W_n$ will involve $\phi_0$ and $W_n^*$ will involve a consistent estimate, $\phi_n$. Similarly, $\phi_n$ will be included in an estimate of $V_n^{*1/2}$. It is also possible to replace $k_g$ with random interval endpoints, $k_{n,g}$, provided that these are consistent estimators of $k_g$ and $P(\beta_0^\top X = k_g) = 0$ for all $g$. This is discussed in more detail in Section \ref{sec:validGLMs}. In our simulation study, described in Section \ref{sec:sim_design}, we use random interval endpoints so that $\sum_{i=1}^n \hat\sigma^2(x_i) I_i^{(g)}$ is approximately equal across groups. Details of this implementation are given in the supplementary material.

Define
\begin{align}
\begin{split}
\label{eq:sigma_n}
    \Sigma_n &= \frac{1}{n} G_n^* \left(V_n^* - V_n^{*1/2} W_n^{1/2} X^* (X^{*\top} W_n X^*)^{-1} X^{*\top} W_n^{1/2} V_n^{*1/2}\right) G_n^{*\top} \\
      &= \frac{1}{n} G_n^* V_n^{*1/2} \left(I_n - W_n^{1/2} X^* (X^{*\top} W_n X^*)^{-1} X^{*\top} W_n^{1/2}\right) V_n^{*1/2} G_n^{*\top} \\
      &= \frac{1}{n} G_n^* V_n^{*1/2} (I_n-H_n) V_n^{*1/2} G_n^{*\top}
\end{split}
\end{align}
where $I_n$ is the $n \times n$ identity matrix, and $H_n = W_n^{1/2} X^* (X^{*\top} W_n X^*)^{-1} X^{*\top} W_n^{1/2}$, also referred to as the generalized hat matrix. We denote the Moore-Penrose pseudoinverse of a matrix $A$ by $A^+$. Our ``generalized HL'' (GHL) test statistic is then given as
\begin{equation}
\label{eq:newtest}
    X^2_{\text{GHL}} = S_n^{1\top} \Sigma_n^{+} S_n^1.
\end{equation}
Under some conditions, described in detail in the Theorem of the Appendix, we have that 
\begin{equation}
\label{eq:conv_in_distbn}
    S_n^{1\top} \Sigma_n^{+} S_n^1 \xrightarrow{d} \chi^2_\nu,
\end{equation}
with $\nu = \rank(\Sigma)$, where $\Sigma$ is specified in the Theorem and supplementary material.

\subsection{GLMs for which the GHL Test is Valid}
\label{sec:validGLMs}
Formally, conditions (A), (B), (B'), (C), and (D) of the Appendix should be verified before using the generalized HL test statistic, $X^2_{\text{GHL}}$. From the Theorem, the conditions below are \textit{sufficient} for the validity of conditions (B), (B'), and (C):  
\begin{enumerate}[(i)]
    \item One of the distribution / link function combinations from Table \ref{tab:distbn_link_combos} is used. (The conditions may hold for other distributions or link functions, but separate verification would be required.)
    \item The joint probability distribution of the explanatory variables, $X$, has compact support, and the distribution of $\beta_0^\top X$ is absolutely continuous. Also, the support of $\beta^\top X$ is an interval for all $\beta$ in a neighbourhood of $\beta_0$.
\end{enumerate}
We reiterate that these are sufficient conditions; our test is not necessarily limited to such GLMs. The supplementary material outlines how to verify conditions (A), (B), (B'), (C), and (D), how to weaken the assumption of compact support from (ii), and how to extend our test to other GLMs. 

It is also generally possible to apply the GHL test to GLMs in which the dispersion parameter associated with the response is unknown by adding appropriate conditions, such as consistency of the estimator of the dispersion parameter. In this case, we have found that a larger sample size is usually needed to ensure that the sampling distribution of the test statistic in finite samples is well approximated by its limiting chi-squared distribution.


\subsection{Related GOF Tests}
\label{sec:related_GOF_tests}
For the simulation study described in Section \ref{sec:sim_design}, we compare our test to tests given by \cite{su1991lack} and \cite{stute2002model}.

Using our notation, the Su-Wei (SW) test statistic is defined as
\begin{equation}
\label{eq:sw_stat}
     X^2_{\text{SW}} = \sup_{u \in \mathbb{R}^d} |\widetilde{R_n}(u)|,
\end{equation}
where $u \in \mathbb{R}^d$,
$$ \widetilde{R_n}(u) = \frac{1}{\sqrt{n}} \sum_{i=1}^n \mathbbm{1}(X_i \leq u) [Y_i - m(\beta_n^\top X_i)],$$
and $\mathbbm{1}(x \leq u)$ is an indicator for the event that each component of $x$ is less than or equal to each respective component of $u$. With continuous covariates, finding the supremum can require $\widetilde{R_n}(u)$ to be evaluated at approximately $n^{d-1}$ values of $u$. Adding to the complexity, obtaining p-values relies on a simulation procedure, described in more detail in \cite{su1991lack}. Even for a relatively small sample size, such as $n=100$, computing the SW test statistic and p-value can be computationally intensive with several predictors.

\cite{stute2002model} present a test statistic based on the Cramér-von Mises statistic applied to a transformed version of the $R_n^1$ process, $T_n R_n^1$. Setting $x_0$ to be, say, the 99th percentile of the observed linear predictors, $\beta_n^\top x_i$, $i=1, \ldots, n$, the Stute-Zhu (SZ) test statistic can be defined as
\begin{equation}
\label{eq:sz_stat}
     X^2_{\text{SZ}} = \frac{1}{n \cdot \psi_n^2(x_0)} \sum_{i=1}^n \mathbbm{1}(\beta_n^\top x_i \leq x_0) [T_n R_n^1(\beta_n^\top x_i)]^2 \cdot \sigma^2_n(\beta_n^\top x_i),
\end{equation}
where
\begin{equation}
\label{eq:psi_SZ_original}
    \psi_n(x_0) = \frac{1}{n} \sum_{i=1}^n \mathbbm{1}(\beta_n^\top x_i \leq x_0) (Y_i - m(\beta_n^\top x_i))^2,
\end{equation}
and $\sigma_n^2(u)$ is a consistent estimator of $\Var(Y|\beta_0^\top x = u)$, satisfying properties mentioned in their paper. The limiting sampling distribution of the test statistic is described in \cite{stute2002model}. 

Although \cite{stute2002model} pose the null hypothesis in terms of the mean only and do not require a distributional assumption in the null hypothesis, we will generally add such an assumption. A distributional assumption places restrictions on the forms of nuisance parameters in the SZ test statistic. For example, in the case of Poisson regression, $\sigma^2_n(u) = e^u$ for all $n$. We also modify the SZ test statistic in the Poisson model by having
$$ \psi_n(x_0) = \frac{1}{n} \sum_{i=1}^n \mathbbm{1}(\beta_n^\top x_i \leq x_0) m(\beta_n^\top x_i).$$
This modification allows the SZ test to detect violations of our narrower null hypothesis, such as overdispersion, that its original formulation would not permit. For the selection of a kernel bandwidth in $T_n R_n^1$, we use a bandwidth of $0.5/\sqrt{n}$, which is used as part of a test in \cite{stute1998model}. 

We denote the percentile of the linear predictors used to define $x_0$ by $p_{x_0}$. We find that calculating $T_n R_n^1$ involves inverting matrices that might not be invertible even when $p_{x_0}$ is much lower than $0.99$, particularly when there are binary covariates or many variables present. We therefore try values of $p_{x_0}$ in $\{0.99, 0.98, 0.95, 0.9, 0.85, 0.8, 0.75, 0.7\}$ in a decreasing order, and use the same value of $p_{x_0}$ across all realizations in a given simulation setting. It is our opinion that having $p_{x_0} < 0.7$ does not include enough of the data for the test statistic to be truly meaningful, and in such cases we omit the calculation of the statistic unless otherwise stated.


\section{Simulation Study Design}
\label{sec:sim_design}
In order to assess the performance of our proposed global GOF test, we perform a simulation study. We focus on applications to data generated from Poisson distributions, since Poisson appears to be a common non-binomial GLM. We refer readers to \cite{surjanovic2021improving} for more information on the performance of the HL test and our generalized HL test in the logistic regression setting. We compare rejection rates under different null and alternative hypothesis settings for four different global GOF tests: the naive generalized HL test, our new generalized HL test (GHL), the SZ test, and the SW test. The naive generalized HL test is included to demonstrate that a test without proper theoretical justification may fail. We believe that the SZ test has an appealing construction, but is perhaps not as well known as the SW test, which can be found in other simulation studies. These two tests are included because they do not rely heavily on kernel-based density estimation and are among the tests that are relatively straightforward to implement. 

Under the null hypothesis we consider six settings with a log link and three with a square root link, varying the distribution of explanatory variables and the true parameter values in each case. To examine the power of each of the GOF tests we consider four different settings representing different model flaws, each with various sub-settings. Finally, we compare the empirical size of tests under the null hypothesis as the number of parameters is increased, keeping the sample size fixed, to illustrate that the naive generalized HL test does not perform well in this setting. Details of all of these cases are given in the next subsections and the supplementary material. 

We use sample sizes of 100 and 500 throughout this simulation study, representing moderate and large sample sizes in many studies in medical and other disciplines where GLMs are used. For each simulation setting we produce 2500 realizations. Only the results for the sample size of 100 are reported for the null and power simulation settings, and important differences between the two settings are summarized in Section \ref{sec:sim_results_application}. Approximate 95\% confidence intervals for the type 1 error rates in the null simulations can be obtained from the observed rejection rates by adding and subtracting $0.01$, whereas conservative 95\% confidence intervals for power can be obtained from the observed rejection rates by adding and subtracting $0.02$, accounting for the widest interval, when a proportion is equal to 0.5. For power comparisons among tests, pairwise McNemar tests are used, along with Bonferroni corrections. For power settings 1, 2, 3, and 4, we account for the 24, 24, 24, and 12 possible comparisons, respectively. We therefore check for p-values less than $2.08 \times 10^{-3}$ or $4.16 \times 10^{-3}$, depending on the power setting. The simulations are performed using \textsf{R}.

\subsection{Null Distribution}
In the first three null settings with a log link, a model with a single covariate and an intercept term is used. These settings serve to examine the effect of small and large fitted values on the null distribution of the test statistics. For these settings, the true regression model is $\Exp(Y|X=x) = \exp(\beta_0 + \beta_1 x)$. The conditional distribution of $Y$ given $X$ is Poisson, and realizations of $X$ are drawn from $U(-3, 3)$, where $U(a,b)$ denotes the uniform distribution on $[a,b]$. The values of $\beta_0$ and $\beta_1$ are chosen so that the fitted values take on a wide range of values (approximately 0.1 to 100) in the first setting, are moderate in size (approximately 1 to 10) for the second setting, and are very small (approximately 0.1 to 1) for the third setting. 

For settings 4, 5, and 6, coefficients are chosen so that the fitted values are moderate in size (rarely less than 1, with an average of approximately 4 or 5), so that other sources of potential problems for the GOF tests can be identified. The fourth setting examines a model including both continuous and dichotomous covariates (the ``Normal-Bernoulli'' model), and is similar to the one used in \cite{hosmer2002goodness}. For this setting the true regression model is
$\Exp(Y|X_1 = x_1, X_2=x_2, B=b) = \exp(\beta_0 + \beta_1 x_1 + \beta_2 x_2 + \beta_3 b)$,
with $B \sim$ Bernoulli$(0.5)$, $(X_1, X_2 | B=0) \sim N((-1, -1), \Sigma)$, and $(X_1, X_2 | B=1) \sim N((1, 1), \Sigma)$, with $\Var(X_1) = \Var(X_2) = 1$, and $\text{Corr}(X_1, X_2) = 0.5$. The fifth and sixth simulation settings examine the effects of correlated and right-skewed covariates, respectively. It is well known that in the presence of multicollinearity the variance of regression parameter estimates can become inflated. The correlated covariates setting is included to assess the impact of multicollinearity on the GOF tests, since the estimated covariance matrix of the regression coefficients is used in the calculation of the GHL test statistic. For this setting, the true regression model is $\Exp(Y|X_1 = x_1, X_2 = x_2) = \exp(\beta_0 + \beta_1 x_1 + \beta_2 x_2)$,
with $(X_1, X_2) \sim N((0, 0), \Sigma)$, so that $\Var(X_1) = \Var(X_2) = 1$, and $\text{Corr}(X_1, X_2) = 0.7$. The right-skewed covariate in setting 6 is included in order to assess its potential impact on the SZ test, since this test makes use of kernel-based density estimation as a part of the calculation of the test statistic, albeit in a one-dimensional case. This setting considers $X \sim \text{Exp}(1)$, where $\text{Exp}(\lambda)$ is the exponential distribution with rate $\lambda$. The true regression model is as in settings 1 to 3. 

The three settings for the square root link Poisson regression simulations are the same as settings 1, 2, and 3 with the log link, except that the true regression model is $\Exp(Y|X=x) = (\beta_0 + \beta_1 x)^2$, and the coefficients $\beta_0$ and $\beta_1$ are modified so that fitted values are approximately within the same three intervals as for settings 1 to 3. All model coefficients and further details of the simulation study are given in the online supplemental materials.

\subsection{Power}
To examine the power of the GOF tests, we consider four types of deviations from the null model: a missing quadratic term, overdispersion, a missing interaction term, and an incorrectly specified link function. These settings are similar to those used in \cite{hosmer2002goodness} and are realistic model misspecifications. In all four settings we choose appropriate regression coefficients so that the fitted values are moderate in size, rarely less than 1 and often smaller than 10 (in setting 3, some fitted values are larger than 10), to ensure that a large rejection rate is not simply due to small fitted values in the Pearson-like test statistics. The size of the deviation from the null model in settings 1 to 3 is indexed by $J$. All four power simulation settings are described in detail in the supplementary material.

\subsection{Performance with Large Models}
\label{sec:large_models}
To assess the performance of each of the tests with large models, for a model with $d$ parameters, we generate data so that $\Exp(Y|X=x) = \exp(\beta_0 + \beta_1 x_1 + \cdots + \beta_{d-1} x_{d-1})$, where $X = (X_1, \ldots, X_{d-1})^\top \sim N(\mathbf{0}, I_{d-1}).$ Realizations of $Y$ are again drawn from a Poisson distribution with the given mean, and we use $n=100$ and $500$ for our sample sizes. In order for the distribution of the fitted values to remain approximately constant as the number of parameters, $d$, is varied, we set
$ \beta_0 = 1.67$, and $\beta_1 = \ldots = \beta_{d-1} = \sqrt{0.0717/(d-1)}$,
resulting in a distribution of fitted values mostly within the interval $[1,10]$, ensuring that expected counts within each group used in the calculation of the Pearson statistic are sufficiently large. We use $d=2, 10, 20, 30, 40, 50$. The SW test is omitted due to computational challenges that arise with this test with large models. Also, the SZ test is omitted since we observe that the choice of $p_{x_0}$ sometimes has to be less than 0.7 when $d$ is large and $n=100$. This is because at least $d-1$ of the matrices that need to be inverted in the calculation of the SZ test statistic have less than full rank.


\section{Simulation Results and Application}
\label{sec:sim_results_application}
\subsection{Simulation Results}
\label{sec:sim_results}

\subsubsection*{Null Distribution}
From the null simulation results in Table \ref{tab:null_results}, we see that the estimated type 1 error rate for our GHL test does not significantly differ from the nominal level, since all values are in the interval $(0.041, 0.059)$. On the other hand, the naive generalized HL test falls out of this interval in three settings, whereas the SW and SZ tests fall out of this interval in two and four settings, respectively. Interestingly, even with $d=4$ in setting 4, we begin to see a decreased type 1 error rate for the naive generalized HL test, a phenomenon discussed in more detail later in this section. However, with a sample size of 500, the naive generalized HL test and the SZ test generally have better empirical rejection rates, whereas the SW test has similar poor performance even with a larger sample size.

\subsubsection*{Power}
\begin{figure}[!htb]
\centering
\begin{subfigure}{.49\textwidth}
  \centering
  \includegraphics[width=\linewidth]{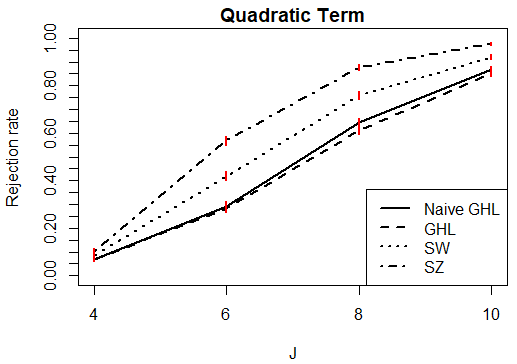}
  \label{fig:sub1}
\end{subfigure}
\begin{subfigure}{.49\textwidth}
  \centering
  \includegraphics[width=\linewidth]{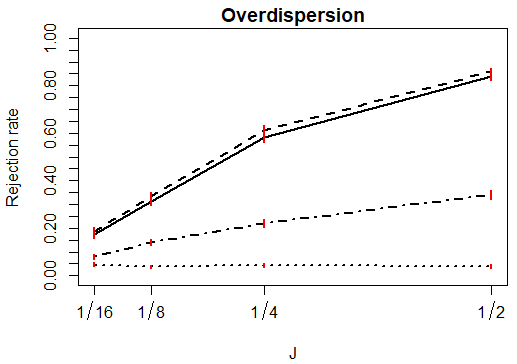}
  \label{fig:sub2}
\end{subfigure}
\begin{subfigure}{.49\textwidth}
  \centering
  \includegraphics[width=\linewidth]{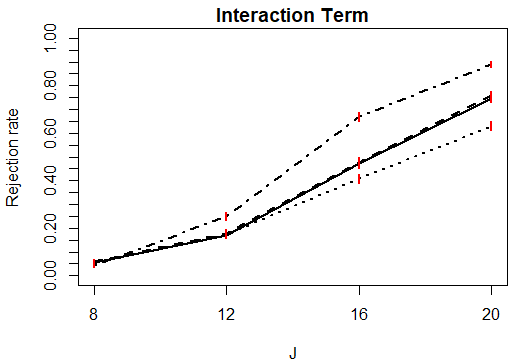}
  \label{fig:sub3}
\end{subfigure}
\caption{Power simulation results for the first three settings. Solid red lines are 95\% Wilson CIs.}
\label{fig:power1to3}
\end{figure}

From the power simulation results displayed in Figure \ref{fig:power1to3}, we see that our new test does, indeed, have power to detect each of the violated model assumptions that we tested. However, for those model flaws that are detectable by the SZ and SW tests, these two tests generally have better power than the tests based on grouped residuals. Using pairwise McNemar tests with a Bonferroni correction, described at the beginning of Section \ref{sec:sim_design}, we see that the SZ test performs significantly better than all of the other tests in the first setting ($p<4.9\times 10^{-6}$ for 11 out of $3 \times 4 = 12$ comparisons), with only the comparison between the SZ and the SW tests at $J=4$ being not significant ($p=8.3 \times 10^{-3}$). In the second setting, we see that our test has the greatest power for detecting overdispersion. The difference is significant in comparison to the naive generalized HL test for $J=1/8, 1/4,1/2$ ($p<9.0 \times 10^{-5}$ for all three comparisons). The difference in power between the GHL test and the SW and SZ tests is significant for all $J$ in this setting ($p<3.9 \times 10^{-34}$ for all 8 comparisons). We note that with the original $\psi_n(x_0)$ from (\ref{eq:psi_SZ_original}) the SZ test is not able to detect overdispersion. Interestingly, in the setting with a missing interaction term, our test performs better than the SW test for $J=16,20$, with the smaller sample size of 100 ($p<4.4 \times 10^{-11}$ for both comparisons). However, with a sample size of 500, not displayed in Figure \ref{fig:power1to3}, the SW test seems to outperform our test. The SZ test has greater power to detect a missing interaction than the GHL test for $J=12, 16, 20$ ($p<7.3 \times 10^{-14}$ for all three comparisons), and the difference between the GHL test and the naive generalized HL test is not significant with this sample size ($p>0.04$ for all four comparisons). The results from the fourth setting are provided in the supplementary material in Table \ref{tab:power4}, but are not displayed here. The SZ test seems to detect an incorrectly specified link function with the most power, although all tests have relatively low power to detect the incorrectly specified link function, possibly due to our choices of alternative link functions, which did not deviate drastically from the fitted model over the range of means chosen. We speculate that global GOF tests should generally be able to detect a misspecified link very well, perhaps better than other model misspecifications, due to trends that can arise in the residuals in such a scenario, provided that the range of means is sufficiently large.

\subsubsection*{Large Models}
\begin{figure}[!htb]
  \begin{center}
    \includegraphics[width=0.65\linewidth]{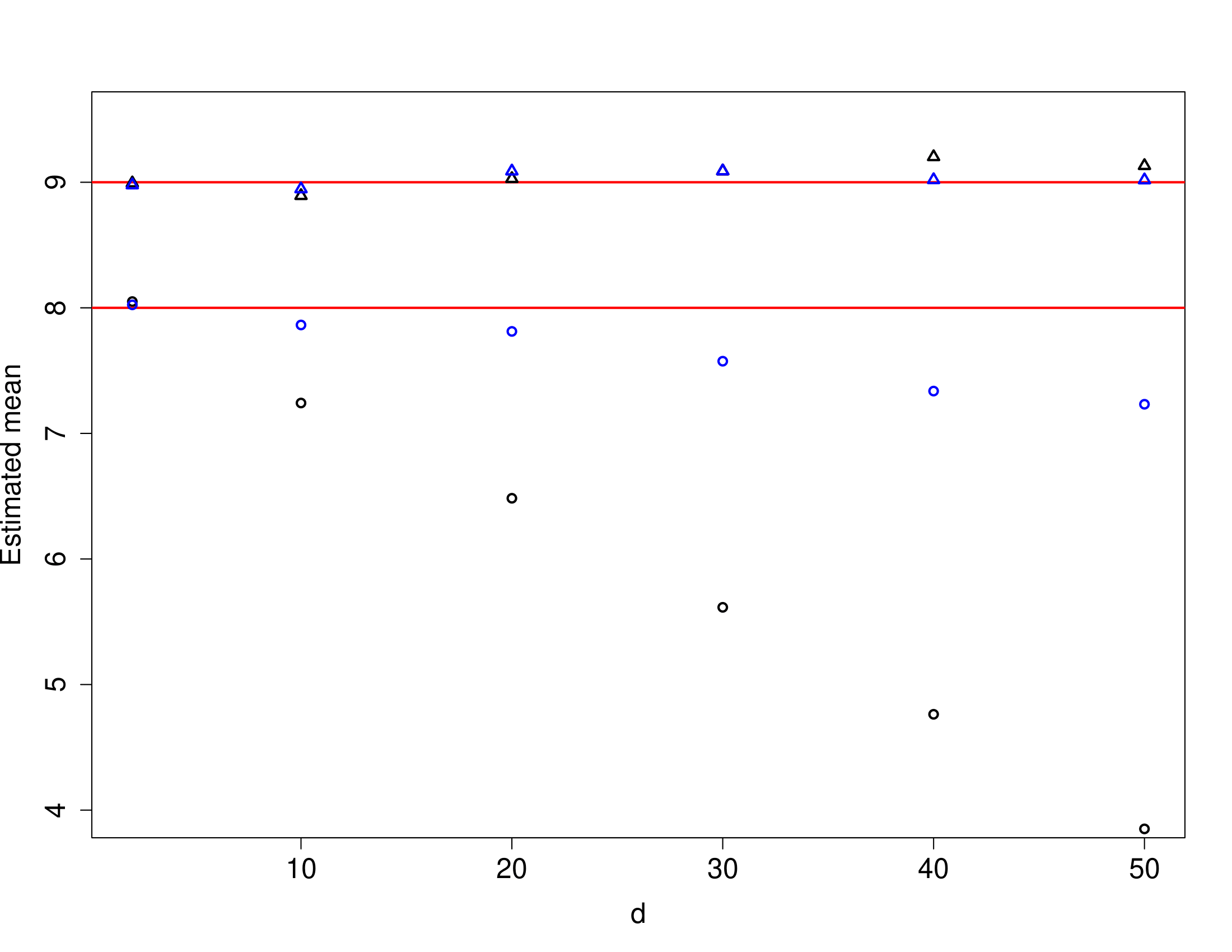}
  \end{center}
  \caption{Estimated mean value of the naive HL (circles) and generalized HL (triangles) statistics for Poisson regression models, using sample sizes of 100 (black) and 500 (blue). Horizontal lines represent the respective true means of the chi-squared distributions to which the test statistics are compared.}
    \label{fig:PoissonHLMeanDecrease}
\end{figure}

The null distribution of the naive generalized HL test statistic, $\widehat{C}_G^*$, is not well-approximated by the usual $\chi^2_{G-2}$ distribution in this particular setting with a finite sample size. The impact of the number of parameters on the estimated mean of the naive generalized HL test statistic can be seen in Figure \ref{fig:PoissonHLMeanDecrease}, where the $G-2$ degrees of freedom approximation for this test deteriorates as the model size grows, relative to the sample size. This adverse effect is less pronounced with a larger sample size. The estimated type 1 error rates are omitted, but they steadily decrease for the naive generalized HL test from about 0.050 to 0.001 as $d$ grows from 1 to 50 with a sample size of 100, and down to 0.030 with a sample size of 500. Our proposed test does not seem to be affected by the number of parameters present in the model. Similar results were obtained for both tests with $G=50$, used to ensure that $G>d$. A parallel simulation was performed in the logistic regression setting, with details given in \cite{surjanovic2021improving}.

\subsection{Application}
\label{sec:application}
We study the alcohol consumption dataset used in the study of \cite{dehart2008drinking}, as described in \cite{bilder2014analysis}. The goal of the study was to assess how the number of alcoholic drinks consumed is associated with factors such as self-esteem and negative romantic-relationship events. \cite{bilder2014analysis} performed a Poisson regression analysis on a subset of the data. The variable NUMALL, representing the number of alcoholic drinks consumed by a subject on their first Saturday, was regressed against several variables. After a fairly extensive analysis of the data, the authors arrived at a final model containing all of the main effects of the variables in Table \ref{tab:alc_vars}, including the following interactions: ROSN $\times$ PREL, AGE $\times$ ROSN, DESIRED $\times$ GENDER, DESIRED $\times$ AGE, and STATE $\times$ NEGEVENT. 

Understanding the questionable validity of performing GOF tests following model selection on the same data, we examine the fit of three different models using the GOF tests discussed in this paper. Additionally, the fit of several other models is examined in the supplementary material. We have $n=89$, and we use $G=10$ and $G=18$ groups. The larger number of groups is included to ensure that $G>d$, as was mentioned in Section \ref{sec:methods}, while still maintaining an average of about 5 observations per group. Because of the relatively high dimension of the data, the SW test is excluded in the evaluations of all three models.

We first examine the overall fit of the model mentioned at the beginning of this subsection (case 1).
As seen in Table \ref{tab:app_results}, none of the tests considered reject the null hypothesis at the $\alpha = 0.05$ significance level. The naive HL gives somewhat larger p-values than the generalized version, matching expectations due to the moderately large number of variables relative to the sample size. For case 2 we fit a square root link instead of the original log link, retaining all of the variables from the model in case 1. The GHL results with both values of $G$ and the naive HL with $G=18$ suggest that the square root link may provide a poorer fit than the log link. However, the naive HL with $G=10$ and the SZ test do not reject the null hypothesis. Finally, in case 3, we omit all interactions from the the model in case 1. Here, we expect that the tests should detect a poor fit due to missing variables. In this case we see that the GHL test with 18 groups, the naive generalized HL test with 10 and 18 groups, and the SZ test reject the null hypothesis. While the GHL test with 10 groups fails to reject the null hypothesis that the model is correct, the p-value is still very close to 0.05.


\section{Discussion}
\label{sec:discussion}
We have introduced several global GOF tests for GLMs and assessed some of their properties. With the simulation results of Section \ref{sec:sim_results_application}, we have seen that our test provides competitive or comparable power in various simulation settings. Our test is also computationally efficient, straightforward to implement, and functions in a variety of scenarios. There is no need for a choice of a kernel bandwidth (although the choice of number of groups, $G$, can play a role in determining the outcome of the test), and the output can be interpreted in a meaningful way by assessing differences between observed and expected counts in each of the groups.

In many of the null and power simulation settings, $d$ was relatively small and the covariates were often continuous. This choice was deliberate, so that power comparisons would be possible, because it is sometimes not possible to calculate the SW and SZ test statistics in the presence of many variables or binary covariates. Also, fewer alternatives are tested by the SW test than by our proposed GHL test. Thus, while SZ and SW were more powerful than the GHL test in certain settings friendly to these tests, there is room for all of these tests in an analyst's toolkit.

The naive generalization of the HL test does not work well under certain settings, but the use of the GHL test seems to resolve these issues. Similar results were found in the logistic regression setting with the regular HL test and the GHL test \citep{surjanovic2021improving}. However, the choice of the number of groups, $G$, can have an impact on the corresponding p-value in group-based tests. As a suggestion, one can choose a value of $G$ so that the values of $\sum_{i=1}^n \hat\sigma^2(x_i) I_i^{(g)}$ or $\sum_{i=1}^n m(\beta_n^\top x_i) I_i^{(g)}$ are larger than, say, 5 in each group, if possible. This is analogous to the common recommendation for the finite-sample validity of the Pearson chi-squared test. Large values of $G$ might be able to increase the power of the GHL test when they can be supported, although this should be studied in more detail.

\section*{Acknowledgements}
We acknowledge the support of the Natural Sciences and Engineering Research Council of Canada (NSERC), [funding reference number RGPIN-2014-06099, RGPIN-2018-04868]. Cette recherche a été financée par le Conseil de recherches en sciences naturelles et en génie du Canada (CRSNG), [numéro de référence RGPIN-2014-06099, RGPIN-2018-04868].

\section*{Supporting Information}
Further information is provided in the ``Supporting Information'' section.

\section*{Appendix}
In this appendix we state conditions (A), (B), (B'), (C), and (D), used in the Theorem. The Theorem describes when the convergence in distribution from (\ref{eq:conv_in_distbn}) holds. Conditions (A), (B), and (C) are borrowed from \cite{stute2002model}, with condition (B) being weakened slightly. We assume the conditional density of $Y$ given $X$ is $f_{Y|X}(y|x) = \exp\left\{y \theta - b(\theta)\right\}$, with respect to some dominating measure, say $\nu$. Let $u$ be the inverse link function that satisfies $\theta_i = u(\beta_0^\top X_i)$. For models without a dispersion parameter, the score function from observation $i$ is $U_i(\beta_0)$, given by $U_i(\beta_0) = X_i u'(\beta_0^\top X_i)\left\{Y_i - m(\beta_0^\top X_i\right)\}$, and the Fisher information matrix is $I_1(\beta_0) = \Exp\left[X X^\top [u'(\beta_0^\top X)]^2 b''\left(u(\beta_0^\top X)\right)\right]
$.

\bigskip
\noindent \textbf{Condition (A)}
\begin{enumerate}[(i)]
    \item $I_1(\beta_0)$ exists and is positive definite.
    \item Let $\ell(X_i,Y_i,\beta_0) = [I_1(\beta_0)]^{-1} U_i(\beta_0)$. Under the null hypothesis, we have that
    $$
    n^{1/2} \{\beta_n - \beta_0\} = n^{-1/2} \sum_{i=1}^n \ell(X_i,Y_i, \beta_0) + o_P(1).
    $$
\end{enumerate}

\bigskip
\vspace{-3mm} \noindent \textbf{Condition (B)} \\
The function $m$ is continuously differentiable on the real line. Define the column vector
    \begin{equation*}
        q(x, \beta) = \frac{\partial m(\beta^\top x)}{\partial \beta} = (q_1(x,\beta), \ldots, q_d(x,\beta))^\top =m'(\beta^\top x) x.
    \end{equation*}
    Letting $\mu$ denote the distribution of $X$, there exists a $\mu$-integrable function $M(x)$ such that, for all $\beta$ in a neighbourhood of $\beta_0$, and for $i = 1, \ldots, d$, $|q_i(x, \beta)| \leq M(x)$. That is, for some $\delta>0$ we have
    $$
    \Exp\left[\sup_{\{\beta: \|\beta-\beta_0\|\le \delta\}} \{\max_j|X_j\ m'(\beta^\top X)| \}\right] < \infty.
    $$
\bigskip
\noindent \textbf{Condition (B')} \\
As before, $\sigma^2(x) = \Var(Y|X=x)$.
\begin{enumerate}[(i)]
    \item The map $(x,\beta) \mapsto \sigma^2(x)$ is continuous.
    
    \item[(ii)] There exists a $\mu$-integrable function $\tilde{M}(x)$ such that, for all $\beta$ in a neighbourhood of $\beta_0$,
    $\sigma^2(x)\leq \tilde{M}(x)$.
\end{enumerate}
\bigskip
\vspace{-3mm} \noindent \textbf{Condition (C)} \\
Define
\begin{equation*}
    \tilde{H}(u, \beta) = \int_{\{ \beta^\top X \leq u\}} 
    \sigma^2(X)\, d\mathbbm{P}=\Exp\left\{\sigma^2(X)1(\beta^\top X \le u)\right\}.
\end{equation*}
$\tilde{H}$ is uniformly continuous in $u$ at $\beta_0$. This condition requires that $\beta_0^\top X$ have a continuous distribution; in particular, $\beta_0 \neq 0$.

\vspace{-3mm} \bigskip \noindent \textbf{Condition (D)} 
\begin{enumerate}[(i)]
    \item $\Sigma_n \xrightarrow{p} \Sigma$, and 
    \item $\text{rank}(\Sigma_n) \xrightarrow{p} \text{rank}(\Sigma)$,
\end{enumerate}
under the null hypothesis. The matrix $\Sigma$ is specified in the supplementary material.

We now state the main theorem of our paper. The proof of this theorem is given in the supplementary material. \vspace{-2mm}
\begin{theorem*}
    Suppose that $\Exp(Y^2) < \infty$ and conditions (A), (B), and (C) hold. Then, under the null hypothesis given by (\ref{eq:H0glm}), with $f_{Y|X}$ as in (\ref{eq:f_y|x}), for non-random $k_g$, $g=0,\ldots,G$ (or random $k_{n,g}$, provided that $k_{n,g}$ is consistent for $k_g$ for each $g$), we have that
    \[ S_n^1 \xrightarrow{d} \text{MVN}_G(0, \Sigma) \equiv S_\infty^1, \]
    where $S_n^1$ is as defined in (\ref{eq:S_n^1}), and $\Sigma$ is given by (\ref{eq:sigma_ggGLM}) and (\ref{eq:sigma_ggprime}) in the supplementary material. Furthermore, if there exists a matrix $\Sigma_n$ that satisfies condition (D), then
    \[ S_n^{1 \top} \Sigma_n^+ S_n^1 \xrightarrow{d} \chi_\nu^2,\]
    with $\nu = \rank(\Sigma)$.
    
   If conditions (i) and (ii) of Section \ref{sec:validGLMs} are satisfied, then conditions (B), (B'), and (C) hold. If (A) additionally holds (consistency and asymptotic normality of the MLE), then the matrix $\Sigma_n$ as given by (\ref{eq:sigma_n}) satisfies condition (D)(i) (consistency for $\Sigma$) under the null hypothesis, for random or fixed interval endpoints (as described above). 
\end{theorem*}

\clearpage
\newpage

\begin{table}
\begin{center}
\begin{tabular}{ll}
\hline
Distribution            & Example possible links \\
\hline 
Normal$(\mu,\sigma^2)$  & identity \\
Bernoulli$(\pi)$        & logit, probit, cauchit, cloglog \\
Poisson$(\lambda)$      & log, square root \\
Gamma($\mu,k$)*         & log \\
IG($\mu,\lambda$)**     & log \\
NB($\mu,k$)***          & log \\
\hline
\end{tabular}
\end{center}
\caption{Several possible distribution and link function combinations. Distributions are parameterized so that $\mu$ or $\pi$ represents the mean of the distribution. *Gamma distribution with variance $\mu^2/k$. **Inverse Gaussian distribution with variance $\mu^3/\lambda$. ***Negative binomial distribution with variance $\mu + \mu^2/k$. Dispersion parameters ($\sigma^2$, $1/k$, $1/\lambda$) are assumed to be known.}
\label{tab:distbn_link_combos}
\end{table}

\begin{table}[ht]
\centering
\begin{tabular}{lccccccccc}
Statistic / Setting & 1 & 1b & 2 & 2b & 3 & 3b & 4 & 5 & 6 \\ \hline
$\widehat{C}_G^*$  & 0.058 & \textit{0.061} & 0.042 & 0.051 & 0.051 & 0.047 & \textit{0.037} & 0.055 & \textit{0.059} \\
$X^2_{\text{GHL}}$ & 0.053 & 0.056 & 0.049 & 0.052 & 0.052 & 0.050 & 0.048 & 0.056 & 0.054 \\
$X^2_{\text{SW}}$  & 0.043 & \textit{0.062} & \textit{0.041} & 0.052 & 0.042 & 0.048 & 0.051 & 0.048 & 0.053 \\
$X^2_{\text{SZ}}$  & \textit{0.032} & 0.045 & \textit{0.039} & \textit{0.039} & 0.044 & \textit{0.041} & --- & 0.045 & 0.049 \\
\hline
\end{tabular}
\caption{Estimated type 1 error rates (null setting simulation results). Numbers in italics represent cases where the estimated rejection rate is significantly different from 0.05 at a significance level of $\alpha=0.05$.}
\label{tab:null_results}
\end{table}

\begin{table}[ht]
\centering
\begin{tabular}{ll}
  \hline
   Variable Name & Description \\ 
  \hline
   NUMALL (response) & Number of alcoholic drinks \\ 
   NEGEVENT & Index for negative events  \\ 
   PREL & Index for romantic-relationship events classified as positive \\ 
   AGE & Age of the subject  \\ 
   ROSN & Long-term (trait) self-esteem level \\ 
   STATE & Short-term (state) self-esteem level \\ 
   GENDER & Gender of the subject \\
   DESIRED & Desire of the subject to drink \\
   \hline
\end{tabular}
\caption{Description of variables used in the alcohol consumption study. Based on the subset of the data used in the Poisson regression analysis of \cite{bilder2014analysis}, each variable is derived from measurements for the first Saturday of each subject in the study.}
\label{tab:alc_vars}
\end{table}

\begin{table}[ht]
\centering
\begin{tabular}{lrrrr}
  \hline
   Statistic & G & Case 1 & Case 2 & Case 3 \\ 
  \hline
   $X^2_{\text{GHL}}$   & 10 & 0.433    & \textit{0.025}& 0.050 \\
   $\widehat{C}_G^*$    & 10 & 0.675    & 0.275         & \textit{0.048} \\
   $X^2_{\text{GHL}}$   & 18 & 0.065    & \textit{0.004}& \textit{0.031} \\
   $\widehat{C}_G^*$    & 18 & 0.118    & \textit{0.035}& \textit{0.029} \\
   $X^2_{\text{SZ}}$    &--- & 0.101   & 0.988        & \textit{0.002} \\
   \hline
\end{tabular}
\caption{GOF test results for various alcohol consumption models. Table C.6 from \cite{klein2006survival} is used to approximate p-values for the SZ test.}
\label{tab:app_results}
\end{table}

\clearpage
\newpage

\begin{center}\textbf{\Large{Supporting Information: A Generalized Hosmer-Lemeshow Goodness-of-Fit Test for a Family of Generalized Linear Models}}\end{center}

\section{Supporting Information Appendix A: Additional Results and Implementation Details}
\subsection{Power Simulation Settings}
\label{sec:power_sim_settings}
In this section we describe the four power simulation settings in greater detail. For the missing quadratic term in setting 1, the true model is 
\[ \Exp(Y|X=x) = \exp(\beta_0 + \beta_1 x + \beta_2 x^2), \]
but we omit the quadratic term when fitting the model. In this setting, we use $X \sim U(-3, 3)$. Four sub-settings are considered, in order to investigate the impact of varying degrees of deviation from the null model. The coefficients $\beta_0, \beta_1, \beta_2$ are chosen so that the mean of $Y$ when $X=-3, 0, 3$ is $J$, $5$, and $8$, respectively. We use $J = 4, 6, 8, 10$. 

In order to simulate overdispersion for setting 2, we draw realizations of $Y$ given $X$ from a negative binomial distribution with mean 
\[ \Exp(Y|X=x) = \exp(\beta_0 + \beta_1 x), \]
and variance
\[  \Var(Y|X=x) = \Exp(Y|X=x) + J \Exp(Y|X=x)^2\] 
for $J = 1/16, 1/8, 1/4, 1/2$. Larger values of $J$ represent larger deviations from the Poisson distribution and greater overdispersion. In this setting we also use $X \sim U(-3, 3)$. 

For setting 3, the true model is 
\[ \Exp(Y|X=x, B=b) = \exp(\beta_0 + \beta_1 x + \beta_2 b + \beta_3 x b), \]
with $X \sim U(-3, 3)$ and $B \sim \text{Bernoulli}(0.5)$. The model that is fit excludes the final interaction term. Four sub-settings are considered for setting 3, similar to setting 1. The coefficients are chosen so that the mean of $Y$ when $(X,B) = (-3,0), (-3, 1), (3, 0), (3,1)$ is equal to 5, 5, 7, and $J$, respectively. We use $J=8, 12, 16, 20$, representing varying amounts of deviation from the fitted model. The SZ test is not omitted in setting 3, even though we were unable to perform some of the simulations even with $p_{x_0} = 0.7$. This can sometimes occur when binary covariates are present in the model because some matrices that should be inverted in the calculation of the test statistic become singular. Instead, data that resulted in an inability to compute the SZ test statistic was omitted, resulting in fewer than 2500 simulation realizations in each sub-setting. 

Finally, in order to assess the effect of an incorrectly specified link function in setting 4, we consider the square root and identity link functions as two sub-settings, fitting a log-linear Poisson model for both. We have $X \sim U(-3, 3)$, and the coefficients are chosen so that the conditional mean of $Y$ given $X=0, 3$ is 5 and 8, respectively.

\subsection{Further Details}
\noindent \textbf{Implementation of the Interval Endpoint Selection Method} \\
For the GHL test, a special interval endpoint selection method is used in order to keep $\sum_{i=1}^n \hat\sigma^2(x_i) I_i^{(g)}$ roughly constant across groups. The implementation is based on the ``weighted.quantile()'' function from the ``spatstat'' package in \textbf{\textsf{R}}. In order to avoid having groups with no observations---this can occur, for example, when a very large fitted value is present---first the weighted $(G-1)/G \times 100$th percentile is obtained. Then, observations that fall into this group are discarded, and the weighted $(G-2)/(G-1) \times 100$th percentile is obtained from the remaining data. This process is repeated until $G$ groups are formed.

\bigskip \noindent \textbf{Degrees of Freedom for the GHL Test} \\
The GHL test statistic is compared to a chi-squared distribution with a certain number of degrees of freedom in order to obtain a p-value. Since the rank of $\Sigma$ is generally unknown, we use the rank of $\Sigma_n$ in its place. The rank of $\Sigma_n$ is very often found to be $G-1$ in the simulation study, with rare exceptions occurring in some simulation realizations (e.g., on some realizations when a categorical covariate is present in the model).

\bigskip \noindent \textbf{GLM Convergence with Noncanonical Links} \\
When fitting a Poisson GLM with identity or square-root link in \textbf{\textsf{R}}, certain issues can arise. In general, we aid convergence of the parameter estimates for non-canonical link models by either providing fitted values from a previous model, or by providing rounded versions of the true parameter values to the ``glm()'' function call. On occasion, fitting the GLM with a noncanonical link results in a warning, in which case the particular simulation realization is omitted, resulting in fewer than 2500 simulation realizations.

\bigskip \noindent \textbf{Alcohol Consumption Study Additional Results} \\
Interestingly, when an outlier is removed from the alcohol consumption dataset and the same model is used as in Case 1, the fit of the model seems to get worse, as suggested by 3/5 of the GOF tests. The fit from the naive GHL test with $G=18$ is only slightly better, but the fit from the SZ test increases reasonably. However, when another variable selection procedure is run without the outlier, the fit seems to become better again for each of the tests. Also, when the variable STATE is additionally removed from the model in Case 3, none of the GOF tests reject the null hypothesis. In this scenario, the naive HL and the GHL tests still have p-values relatively close to 0.05.


\subsection{Assorted Tables}
\label{sec:assorted_tables}
\begin{table}[!htb]
\begin{center}
\begin{tabular}{lccc}
  \hline
  Link function name & Link function form & Inverse link function form \\
  \hline
  identity & $\mu$ & $\beta^\top x$ \Tstrut\\
  log & $\log(\mu)$ & $\exp(\beta^\top x)$\\
  logit & $\logit(\mu) = \log(\mu/(1-\mu))$ & $\exp(\beta^\top x)/(1+\exp(\beta^\top x))$\\
  probit & $\probit(\mu) = \Phi^{-1}(\mu)$ & $\Phi(\beta^\top x)$\\
  cauchit & $\cauchit(\mu) = \tan(\pi(\mu - 1/2))$ & $1/\pi \arctan(\beta^\top x) + 1/2$\\
  cloglog & $\cloglog(\mu) = \log(-\log(1-\mu))$ & $1-\exp(-\exp(\beta^\top x))$\\
  square root & $\sqrt{\mu}$ & $(\beta^\top x)^2$\\
  \hline
\end{tabular}
\end{center}
\caption{Link and inverse link functions considered}
\label{tab:linkfunctions}
\end{table}

\begin{table}[!htb]
\begin{center}
\begin{tabular}{lccccc}
\hline
Distribution & $\theta$ & $\Theta$ & $b(\theta)$ & $v(m)$ \\
\hline 
Normal$(\mu,\sigma^2)$ & $\mu/\sigma^2$ & $(-\infty,\infty)$ 
& $\theta^2 \sigma^2/2$ & $\sigma^2$
\\
Bernoulli$(\pi)$ & $\log(\pi/(1-\pi))$ & $(-\infty,\infty)$ 
& $\log(1+e^\theta)$ & $m(1-m)$
\\
Poisson$(\lambda)$ & $\log(\lambda)$ & $(-\infty,\infty)$ 
& $e^\theta$ & $m$
\\
Gamma($\mu,k$) & $-k/\mu$ & $(-\infty, 0)$
& $-k\log(-\theta)$ & $m^2/k$
\\
IG($\mu,\lambda$)* & $-\lambda/(2\mu^2)$ & $(-\infty,0)$ 
& $-\sqrt{-2\theta \lambda}$ & $m^3/\lambda$
\\
NB($\mu,k$)** & $\log(\mu/(\mu+k))$ & $(-\infty, 0)$ 
& $-k\log(1-e^\theta)$ & $m+m^2/k$ \\
\hline
\end{tabular}
\end{center}
\caption{Several conditional distributions of $Y$ given $X$ that can be written in the form of (\ref{eq:f_y|x_phi}).}
\label{tab:distributions}
\end{table}

\begin{table}[!htb]
\centering
\begin{tabular}{lcc}
  \hline
   Setting & Distribution of Covariate(s) & True coefficients \\ 
  \hline
   1 & $X \sim U(-3, 3)$ & $\beta_0 = 1.15, \beta_1 = 1.15$ \\ 
   1b & $X \sim U(-3, 3)$  & $\beta_0 = 5.16, \beta_1 = 1.61$ \\ 
   2 & $X \sim U(-3, 3)$ & $\beta_0 = 1.15, \beta_1 = 0.384$ \\ 
   2b & $X \sim U(-3, 3)$ & $\beta_0 = 2.08, \beta_1 = 0.360$ \\ 
   3 & $X \sim U(-3, 3)$ & $\beta_0 = -1.15, \beta_1 = 0.384$ \\ 
   3b & $X \sim U(-3, 3)$ & $\beta_0 = 0.658, \beta_1 = 0.114$ \\ 
   4 & Normal-Bernoulli model & $\beta_0 = 1, \beta_1 = 0.2, \beta_2 = -0.2, \beta_3 = 0.7$ \\ 
   5 & Correlated covariates & $\beta_0 = 1.70, \beta_1 = 0.148, \beta_2 = 0.148$ \\ 
   6 & $X \sim \text{Exp}(1)$ & $\beta_0 = 1.15, \beta_1 = 0.384$ \\ 
   \hline
\end{tabular}
\caption{Null simulation settings}
\label{tab:null_settings}
\end{table}

\begin{table}[!htb]
\centering
\begin{tabular}{lcc}
  \hline
   Setting & Description & True coefficients \\ 
  \hline
   1 & Missing quadratic term  & \makecell{$\beta_0 = 1.61$, \, $\beta_1 = 0.347 - 1/6 \log(J)$, \\ $\beta_2 = -0.0633 + 1/18 \log(J)$} \\
   2 & Overdispersion & $\beta_0 = 1.61, \beta_1 = 0.157$ \\ 
   3 & Missing interaction term & \makecell{$\beta_0 = 1.78$, \, $\beta_1 = 0.0561$, \, $\beta_2 = 1/2 \log(J/5)$, \\ $\beta_3 = 1/6 \log(J/5)$} \\ 
   4 & Incorrectly specified link & \makecell{$\beta_0 = 2.24, \, \beta_1 = 0.197$ (square root) \\ $\beta_0 = 5, \, \beta_1 = 1$ (identity)}  \\
   \hline
\end{tabular}
\caption{Power simulation settings}
\label{tab:log_Poisson_power_settings}
\end{table}

\begin{table}[!htb]
\centering
\begin{tabular}{lcc} 
Statistic / Link & Square root & Identity \\ \hline
$\widehat{C}_G^*$  & 0.063 & 0.108 \\ 
$X^2_{\text{GHL}}$ & 0.062 & 0.108 \\ 
$X^2_{\text{SW}}$  & 0.054 & 0.130 \\ 
$X^2_{\text{SZ}}$  & 0.057 & 0.136 \\ 
\hline
\end{tabular}
\caption{Power simulation results - incorrect link function}
\label{tab:power4}
\end{table}

\clearpage
\newpage

\section{Supporting Information Appendix B: Discussion and Proof of the Theorem}
\label{sec:discussion_of_theorem}
We primarily concern ourselves with the latter half of the Theorem. The first half of the Theorem is implicitly proven in Section~\ref{sec:obtaining_estimator_of_Sigma} below. We show that conditions (B), (B'), and (C) are satisfied for GLMs that satisfy \textit{both} conditions (i) and (ii), mentioned in Section \ref{sec:validGLMs} of the paper. The results of this paper, however, are not necessarily limited to such GLMs. We also offer general discussion for conditions (A), (B), (B'), (C), and (D). Condition (A) generally holds for maximum likelihood estimators in both standard GLMs and in models with a dispersion parameter. We also provide some comments on condition (D), as well as alternative approaches that can be taken should this condition not be met. Later in the supplementary material we show that the matrix $\Sigma_n$ as given by (\ref{eq:sigma_n}) in the paper satisfies condition (D)(i) under the null hypothesis and conditions (i) and (ii).

We begin by restating the model under our null hypothesis. The sequence $(X_1,Y_1),\ldots$ is iid. For a GLM we assume the conditional density of $Y$ given $X$ is
$$
f_{Y|X}(y|x) = \exp\left\{y \theta - b(\theta)\right\},
$$
with respect to some dominating measure, say $\nu$.
The mean of $Y$ (conditioned on $X$) is $b'(\theta)$ and the variance of $Y$ is $b''(\theta)$. The mean of $Y$ is related to $x$ via $b'(\theta) = m(\beta_0^\top x)$ for some $\beta_0$ and a specified function $m$. Since $b''(\theta)>0$, the function $b'$ is injective on the natural domain $\Theta =\{\theta: \int \exp(y\theta) \nu(dy)<\infty\}$. This means that we may write $\theta$ as a function of $m(\beta_0^\top x)$.

Let $u$ be the inverse link function that satisfies
$$
\theta_i = u(\beta_0^\top X_i).
$$
(This is the composition of the inverse function of $b'$ and the mean function $m$. Note that $u$ is the identity function if $m$ is the canonical link.) For models without a dispersion parameter, the score function from observation $i$ is $U_i(\beta_0)$, given by
$$
U_i(\beta_0) = X_i u'(\beta_0^\top X_i)\left\{Y_i - m(\beta_0^\top X_i\right)\},
$$
and the Fisher information matrix is
$$
I_1(\beta_0) = \Exp\left[X X^\top [u'(\beta_0^\top X)]^2 b''\left(u(\beta_0^\top X)\right)\right].
$$

\subsection{Verifying Conditions (A), (B), (B'), (C), and (D) for Various GLMs}
\subsubsection*{Condition (A)}
Conditions for asymptotic normality and weak consistency of the MLE $\beta_n$ for $\beta_0$ in GLMs can be found in \cite{fahrmeir1985consistency}. They show that in a GLM \textit{without} a dispersion parameter and with the canonical link function there is, under very mild conditions, a root $\beta_n$ of the likelihood equations that is consistent for $\beta_0$ in the case of iid covariates.

We first consider canonical (natural) link functions, since such links make the log-likelihood convex. If $X$ has compact support and $\Exp(X X^\top)$ is positive definite (i.e., the covariance matrix of $X$ is positive definite), then $\beta_n$ asymptotically exists and is strongly consistent, by Corollary 3 of \cite{fahrmeir1985consistency}. We now consider the case where the support of $X$ is not compact and introduce condition $(R_s)$ of \cite{fahrmeir1985consistency}:
\begin{itemize}
    \item[] $(R_s) (i): I_1(\beta_0)$ exists and is positive definite, and
    \item[] $(R_s) (ii): \Exp\left[ \max_{\beta \in N} X X^\top b''\left(\beta^\top X\right)\right]$ exists for some compact neighbourhood, $N$, of $\beta_0$.
\end{itemize}
If condition $(R_s)$ holds, then $\beta_n$ asymptotically exists and is strongly consistent, also by their Corollary 3. The expansion in condition (A) follows. 

For link functions other than the canonical link, it can be the case that the score function has multiple roots. In this case, results from \cite{fahrmeir1985consistency, fahrmeir1986correction} assert that there is a consistent root and that this root satisfies the expansion in condition (A). Some additional conditions might be required to establish weak consistency of $\beta_n$ with noncanonical links.

\subsubsection*{Condition (B)}
The inverse link functions considered are mentioned in Section \ref{sec:validGLMs} of the paper. All of these are continuously differentiable on the whole real line, so the first part of condition (B) is met. 

The second part of condition (B) imposes moment conditions on the covariates.  Consider first the log link, for which $m(\mu)=e^\mu$. If there is a neighbourhood of $\beta_0$ in which $X$ has a moment generating function then condition (B) holds. For the logit, probit, cauchit, cloglog, and identity links it is enough that $X$ have a finite mean; that is, for each $1 \le j \le d$ we have $\Exp(|X_j|) < \infty$. For the square root link we require finite variances: for each $j$ we require $\Exp(X_j^2) < \infty$. All of these moment conditions are immediate consequences of an overall assumption that the covariates $X$ lie in some bounded set with probability 1. Here are some details.

Consider first the log link, i.e.,~$m(\beta^\top x) = \exp(\beta^\top x)$. Suppose that for some $\epsilon>0$ the random vector $X$ has a finite moment generating function $\Exp(\exp(\beta^\top X))<\infty$ for all $\beta$ such that $\|\beta-\beta_0\| \le \epsilon$. Then it can be shown using the convexity of $\beta\mapsto \exp(\beta^\top x)$ that for any $0<r<\epsilon$  and any finite $\tau$ we have
$$
\Exp\left[ \sup_{\{\beta:\|\beta-\beta_0\| \le r\}} \|X\|^\tau \exp\{\beta^\top X\}\right]<\infty.
$$
Condition (B) then holds with
$$M(x) = \|x\| \exp\left\{(\|\beta_0\| + r) \|x\|\right\}.$$ 
Indeed, for $i = 1,\ldots, d$, we have, for all $\beta$ with $\|\beta-\beta_0\|\le r$,
\begin{align*}
        |q_i(x, \beta)| &= \left \lvert
        m'\left(\beta^\top x\right) x_i
        \right \rvert \\
        &= \left \lvert x_i \exp(\beta^\top x) \right \rvert\\
        &\leq M(x) .
\end{align*}

For the identity link we have $q_i(x,\beta) = x_i$. For the four bounded links the derivative $m'$ is also bounded and we have
$$
| q_i(x,\beta)|\le \|x\| \|m'\|_\infty.
$$

For the square root link we see $m'(\mu)=2\mu$ and 
$$
| q_i(x,\beta)
|\le 2 \|x\| \sup_{\{\beta:\|\beta-\beta_0\| \le r\}}|\beta^\top x| 
\le 2 \|x\|^2 \sup_{\{\beta:\|\beta-\beta_0\| \le r\}}\|\beta\|^2.
$$

\subsubsection*{Condition (B')}

Let $\Theta$ be the set of $\theta$ for which $\int\exp(y\theta)\nu(dy)<\infty$. Then $\Theta$ is an interval in the real line. On the interior of that interval the function $b$ has infinitely many derivatives so in particular $b$, $b'$ and $b''$ are all continuous on the interior of $\Theta$. Continuity of the map $(x, \beta) \mapsto \sigma^2(x)$ holds, since 
\[ \sigma^2(x) = b''(\theta) = b''(b'^{-1}(m(\beta^\top x)).\]
All of the inverse link functions considered from Section \ref{sec:validGLMs} are continuous on the real line, and continuity is preserved when continuous functions are composed, multiplied, or added. All $\tilde{M}(x)$ considered below are $\mu$-integrable, since we have assumed that the support of the joint probability distribution of the explanatory variables is compact. Weaker moment conditions like those discussed under Condition (B) above are straightforward, but the details depend on the specific variance function and the specific link; some details are provided below in a case by case list of the models. \vspace{2mm} \\ 
\textbf{Normal($\mu, \sigma^2$)} \\
Here, $\sigma^2(x) = \sigma^2$. Then $\tilde{M}(x) = \sigma^2$, is a $\mu$-integrable function such that 
$$\left|\sigma^2(x) \right| = \sigma^2 = \tilde{M}(x), \quad \text{for all $\beta$ in a neighbourhood of $\beta_0$.}$$ 
\textbf{Bernoulli($\pi$)}\\
For the Bernoulli distribution, $\left|\sigma^2(x) \right| = |m^*(x)(1-m^*(x))| \leq 1/4 $, and we set $\tilde{M}(x)=1/4$. \vspace{2mm} \\
\textbf{Poisson($\lambda$)} \\
For the Poisson distribution, $\left|\sigma^2(x)\right| = |m^*(x)|$. If, for instance, $m^*(x) = \exp(\beta^\top x)$ and $\|\beta - \beta_0\| \leq r$, then
\begin{align*}
    |m^*(x)| &= \left| \exp(\beta^\top x) \right| \\
             &\leq \exp(\|\beta\| \|x\|) \\
             &\leq \exp((\|\beta_0\| + r) \|x\|).
\end{align*}
We can then set $\tilde{M}(x) =  \exp((\|\beta_0\| + r) \|x\|)$. On the other hand, with the square root link, 
\begin{align*}
    |m^*(x)| &= \left| (\beta^\top x)^2 \right| \\
             &\leq \|\beta\|^2 \|x\|^2 \\
             &\leq (\|\beta_0\| + r)^2 \|x\|^2.
\end{align*}
Thus for this model we get the same moment conditions for these two links as we had when discussing Condition (B). Other link functions can be handled similarly. \vspace{2mm} \\
\textbf{Gamma($\mu$, $k$), IG($\mu, \lambda$), NB($\mu,k$)} \\
For the gamma distribution with mean $m^*(x)$ and shape parameter $k>0$, $\sigma^2(x) = m^*(x)^2/k $. For the inverse Gaussian distribution with mean $m^*(x)$ and shape parameter $\lambda>0$, we have that $\sigma^2(x) = m^*(x)^3/\lambda$. For the negative binomial distribution with mean $m^*(x)$ and known parameter $k$, $\sigma^2(x) = m^*(x) + m^*(x)^2/k$. For any one of these, the approach is similar to the one taken in the Poisson case.

\subsubsection*{Condition (C)}
This condition requires that $\beta_0^\top X$ have a continuous distribution. In particular, it will not be satisfied with only discrete covariates or with $\beta_0=0$. This is a real restriction; if the covariates are unrelated to the response then in the limit all of the observations will be in a single cell.

\subsubsection*{Condition (D)}
We later show that if conditions (i), (ii), and (A) hold, then for $\Sigma_n$ given by (\ref{eq:sigma_n}), $\Sigma_n \xrightarrow{p} \Sigma$, under the null hypothesis. Therefore, condition (D)(i) is satisfied. However, condition (D)(ii), that $\text{rank}(\Sigma_n) \xrightarrow{p} \text{rank}(\Sigma)$, can be more difficult to verify. Our simulation results from Section \ref{sec:sim_results_application} suggest that the verification of condition (D)(ii) should not be a major concern. Nevertheless, we provide an alternative approach that can be taken to avoid this potential problem.

Along the lines of Proposition 2 of \cite{lutkepohl1997modified} and the ``trimmed'' or ``winsorized'' tests of \cite{davidov2018testing}, the main idea is to make use of the eigendecompositions of $\Sigma_n$ and $\Sigma$, so that $\Sigma_n = E_n \Lambda_n E_n^\top$ and $\Sigma = E \Lambda E^\top$, where the columns of $E, E_n$ are orthogonal, and $\Lambda, \Lambda_n$ are diagonal matrices. Then, we can ``trim'' $\Sigma_n$ by setting all entries of $\Lambda_n$ that are smaller than some $\epsilon > 0$ to zero. This prevents undesirable instabilities when making use of generalized inverses of $\Sigma_n$ in test statistics. We refer readers to \cite{lutkepohl1997modified} and \cite{davidov2018testing} for more information.

\subsection{Obtaining an Estimator of $\Sigma$}
\label{sec:obtaining_estimator_of_Sigma}
We obtain a consistent estimator of $\Sigma$. Throughout, we assume that conditions (i), (ii), and (A) are satisfied. This implies that conditions (B), (B'), and (C) are also met. In the following discussion the interval endpoints, $k_g$, are fixed, although the reasoning extends to the case where random interval endpoints are used, as described in the Theorem. It is also assumed that $\Exp(Y^2) < \infty$, $f_{Y|X}$ is as in (\ref{eq:f_y|x}), and that the null hypothesis is true.

\bigskip \noindent \textbf{Further Notation} \\
We borrow and modify some of the notation used in \cite{stute2002model}. Under our conditions they show that the sequence of processes $R_n^1$ converges weakly to a centered Gaussian process $R_\infty^1$ with the structure
$$
R_\infty^1(u)= R_\infty(u) - Q^\top(u) \Gamma,
$$
whose terms we now describe.
The process $R_\infty$ is a centered Gaussian process and has covariance kernel $K(s, t) = \psi(\min \{s,t\})$, where
$$ \psi(x) = \int_{-\infty}^x \Var(Y | \beta_0^\top X = u) F_{\beta_0} (du), $$
and  $F_{\beta_0}$ denotes the distribution of $\beta_0^\top X$. The vector valued function 
$ Q^\top = (Q_1, \ldots, Q_d)$, is given by
$$ 
Q_i(u) \equiv Q_i(u, \beta_0) = \Exp(q_i(X, \beta_0) \mathbbm{1}(\beta_0^\top X \leq u)). $$
Finally, $\Gamma$ is a $d$-dimensional normal vector with zero means and covariance matrix $[I_1(\beta_0)]^{-1}$, the inverse of the Fisher information matrix for a single observation. From \cite{stute2002model},
$$ 
\Cov(R_\infty(u), Q^\top(s,\beta_0)\Gamma) = Q^\top(s,\beta_0) 
\Exp(\mathbbm{1}(\beta_0^\top X \leq u) \, \epsilon \, \ell(X,Y,\beta_0)). $$
For fixed interval endpoints, $k_g$, that do not depend on the observed data, we therefore have
$$ S_n^1 \xrightarrow{d} \left\lbrace (R_\infty(k_g) - Q^\top(k_g) \Gamma) - (R_\infty(k_{g-1}) - Q^\top(k_{g-1}) \Gamma)\right\rbrace_{g=1}^G \equiv S_\infty^1, $$
where $S_\infty^1$ is multivariate normal, $ \text{MVN}_G(0, \Sigma)$. We now specify the elements of the matrix $\Sigma$.

\bigskip \noindent \textbf{Elements of the covariance matrix $\Sigma$} \\
We first look at the diagonal elements of $\Sigma$.
\begin{align}
\label{eq:sigma_gg}
\Sigma_{gg} &= \Var(R_\infty(k_g) - R_\infty(k_{g-1})) \notag \\
&\qquad + \Var((Q^\top(k_g) - Q^\top(k_{g-1}))\Gamma) \notag \\
&\qquad - 2 \, \Cov(R_\infty(k_g) - R_\infty(k_{g-1}), (Q^\top(k_g) - Q^\top(k_{g-1}))\Gamma) \\
&= \int_{k_{g-1}}^{k_g} \Var(Y|\beta_0^\top X = u) F_{\beta_0}(du) \notag \\
&\qquad + (Q(k_g) - Q(k_{g-1}))^\top [I_1(\beta_0)]^{-1} (Q(k_g) - Q(k_{g-1})) \notag \\
&\qquad -2 \left( (Q(k_g) - Q(k_{g-1}))^\top \Exp\left(\mathbbm{1}(k_{g-1} < \beta_0^\top X \leq k_g) \, \epsilon \, \ell(X,Y,\beta_0)\right) \right) \nonumber \\
\label{eq:sigma_ggGLM}
&= \int_{k_{g-1}}^{k_g} \Var(Y|\beta_0^\top X = u) F_{\beta_0}(du) - \left\lbrace (Q(k_g) - Q(k_{g-1}))^\top [I_1(\beta_0)]^{-1} (Q(k_g) - Q(k_{g-1}))\right\rbrace ,
\end{align}
because it can be shown that 
\begin{equation}
  \label{eq:expectationidentity}
    [I_1(\beta_0)]^{-1} (Q(k_g) - Q(k_{g-1})) = \Exp\left(\mathbbm{1}(k_{g-1} < \beta_0^\top X \leq k_g) \, \epsilon \, \ell(X,Y,\beta_0)\right).
\end{equation}
We now show that the above equation (\ref{eq:expectationidentity}) holds. From \cite{agresti2015foundations}, for the models under consideration, 
$$ U_i(\beta_0) = \left( \frac{(Y-m(\beta^\top X_i))X_i}{\sigma^2(X_i)} \right) m'(\beta^\top X_i)\Biggr\rvert_{\beta = \beta_0}.$$
Therefore,
\begin{align*}
\Exp\left(\mathbbm{1}(k_{g-1} < \beta^\top_0 X \leq k_g) \, \epsilon \, \ell(X,Y,\beta_0)\right) &= [I_1(\beta_0)]^{-1} \cdot \Exp\left( \left( \frac{(Y-m(\beta^\top X))^2}{\sigma^2(X)} \right) m'(\beta^\top X) X \mathbbm{1}(k_{g-1} < \beta^\top X \leq k_g) \right) \Biggr\rvert_{\beta = \beta_0} \\
&= [I_1(\beta_0)]^{-1} \Exp\left( m'(\beta^\top X) X \mathbbm{1}(k_{g-1} < \beta^\top X \leq k_g) \right)  \Biggr\rvert_{\beta = \beta_0}\\
&= [I_1(\beta_0)]^{-1} \Exp\left( q(X, \beta) \mathbbm{1}(k_{g-1} < \beta^\top X \leq k_g) \right)  \Biggr\rvert_{\beta = \beta_0}\\
&= [I_1(\beta_0)]^{-1} (Q(k_g) - Q(k_{g-1})).
\end{align*}

Next, for $g \neq g'$,
\begin{align*}
\Sigma_{gg'} &= \Cov(R_\infty(k_g) - R_\infty(k_{g-1}), R_\infty(k_{g'}) - R_\infty(k_{g'-1})) \\
&\qquad + \Cov((Q^\top(k_g) - Q^\top(k_{g-1}))\Gamma, (Q^\top(k_{g'}) - Q^\top(k_{g'-1}))\Gamma) \\
&\qquad - \Cov(R_\infty(k_g) - R_\infty(k_{g-1}), (Q^\top(k_{g'}) - Q^\top(k_{g'-1}))\Gamma) \\
&\qquad - \Cov((Q^\top(k_g) - Q^\top(k_{g-1}))\Gamma, R_\infty(k_{g'}) - R_\infty(k_{g'-1})).
\end{align*}
However, 
$$ \Cov(R_\infty(k_g) - R_\infty(k_{g-1}), R_\infty(k_{g'}) - R_\infty(k_{g'-1})) = 0,$$
from the covariance kernel of $R_\infty$. Also, the last three terms are equal in absolute value, because
\begin{align*}
  \Cov((Q^\top(k_g) - Q^\top(k_{g-1}))\Gamma,& (Q^\top(k_{g'}) - Q^\top(k_{g'-1}))\Gamma) 
  \\&= (Q(k_g) - Q(k_{g-1}))^\top \Cov(\Gamma,\Gamma) (Q(k_{g'}) - Q(k_{g'-1})) \\
  &= (Q(k_g) - Q(k_{g-1}))^\top [I_1(\beta_0)]^{-1} (Q(k_{g'}) - Q(k_{g'-1})),
\end{align*}
and
\begin{align*}
  \Cov(R_\infty(k_g) - R_\infty(k_{g-1}),& (Q^\top(k_{g'}) - Q^\top(k_{g'-1}))\Gamma) 
  \\&= \Cov(R_\infty(k_g) - R_\infty(k_{g-1}), Q(k_{g'})^\top \Gamma) \\
   &\qquad - \Cov(R_\infty(k_g)- R_\infty(k_{g-1}), Q(k_{g'-1})^\top \Gamma) \\
 &= Q(k_{g'})^\top \Exp\left(\mathbbm{1}(k_{g-1} < \beta_0^\top X \leq k_g) \, \epsilon \, \ell(X,Y,\beta_0)\right) \\
   &\qquad - Q(k_{g'-1})^\top \Exp\left(\mathbbm{1}(k_{g-1} < \beta_0^\top X \leq k_g) \, \epsilon \, \ell(X,Y,\beta_0)\right) \\
 &= (Q(k_g) - Q(k_{g-1}))^\top [I_1(\beta_0)]^{-1} (Q(k_{g'}) - Q(k_{g'-1})),
\end{align*}
by making use of equation (\ref{eq:expectationidentity}) from above. Similarly,
\begin{multline*}
 \Cov\left((Q^\top(k_g) - Q^\top(k_{g-1}))\Gamma, R_\infty(k_{g'}) - R_\infty(k_{g'-1})\right) \\= (Q(k_g) - Q(k_{g-1}))^\top [I_1(\beta_0)]^{-1} (Q(k_{g'}) - Q(k_{g'-1})).
\end{multline*}
Therefore,
\begin{equation}
\label{eq:sigma_ggprime}
    \Sigma_{gg'} = - (Q(k_g) - Q(k_{g-1}))^\top [I_1(\beta_0)]^{-1} (Q(k_{g'}) - Q(k_{g'-1})).
\end{equation}

The proof of the consistency of $\Sigma_n$, as defined in (\ref{eq:sigma_n}) of the paper, is given below. In particular, if conditions (i) and (ii) of Section~\ref{sec:validGLMs} hold, and if condition (A) additionally holds, then 
$$
\Sigma_n = \frac{1}{n} G_n^* (V_n^* - V_n^{*1/2} W_n^{1/2} X^* (X^{*\top} W_n X^*)^{-1} X^{*\top} W_n^{1/2} V_n^{*1/2}) G_n^{*\top} \xrightarrow{p} \Sigma,
$$
under the null hypothesis. This result holds for fixed interval endpoints, $k_g$, or random interval endpoints, $k_{n,g}$, provided that the latter are consistent for $k_g$ and $P(\beta_0^\top X = k_g) = 0$ for all $g$. (Condition (C) guarantees this latter condition.)

Finally, under condition (D)(ii) and by Theorem 1 of \cite{andrews1987asymptotic}, 
$$S_n^{1\top} \Sigma_n^{+} S_n^1 \xrightarrow{d} \chi^2_\nu, $$
where $\nu = \rank(\Sigma)$.

\subsection{Proof of Consistency of $\Sigma_n$}
We show that $\Sigma_n$, defined in (11), is a consistent estimator of $\Sigma$ under some conditions. Define
\begin{align*}
w_A(u) & = \frac{d}{du} v(m(u)) = m'(u) v'(m(u)),
\\
w_B(u) &= m''(u), \text{ and}
\\
w_C(u) &= \frac{d}{du} \frac{\left(m'(u)\right)^2}{v(m(u))}.
\end{align*}
Then, for $\delta>0$, define
\begin{align*}
M_{A,\delta}(x) &= \|x\|\sup\{|w_A(\beta^\top x )|,\|\beta-\beta_0\|\le \delta\},
\\
M_{B,\delta}(x) & = \|x\|^2\sup\{|w_B(\beta^\top x)|,\|\beta-\beta_0\|\le \delta\}, \text{ and}
\\
M_{C,\delta}(x) &= \|x\|^3 \sup\left\{|w_C(\beta^\top x)|,\|\beta-\beta_0\|\le \delta\right\}.
\end{align*}
We now introduce another condition that will help prove the consistency of $\Sigma_n$. 

\bigskip
\noindent\textbf{Condition (E)}\\
There is a $\delta>0$ such that 
$$
\Exp\left\{M_{A,\delta}^2(X)\right\} <\infty,
$$
$$
\Exp\left\{M_{B,\delta}^2(X)\right\} <\infty,
$$
and
$$
\Exp\left\{M_{C,\delta}(X)\right\} <\infty.
$$
Moreover:
$$
\Exp\left[v^2(m(\beta_0^\top X))\right]< \infty,
$$
$$
\Exp\left[\left\{m'(\beta_0^\top X)\right\}^2\|X\|^2\right]< \infty,
$$
and
$$ 
\Exp\left[\frac{\left\{m'(\beta_0^\top X)\right\}^2}{v(m(\beta_0^\top X))} \|X\|^2 \right] < \infty. 
$$
\textbf{Result:} Under conditions (i), (A), (B), (B'), (C), and (E), our estimator $\Sigma_n$ is consistent for $\Sigma$ under the null hypothesis. That is, condition (D)(i) is satisfied.

\noindent \textbf{Remark:} As a consequence of this result, for the GLMs listed in Section 3.4, it is not necessary for $X$ to have compact support as in condition (ii). Provided that (i), (A), (B), (B'), and (C) hold, along with the moment conditions listed in Table \ref{tab:moment_conditions}, our estimator $\Sigma_n$ is consistent for $\Sigma$ under the null hypothesis.  

\begin{table}[ht]
\begin{center}
\begin{tabular}{lccl}
\hline
Distribution            & $m(u)$    & $v(m)$        & Moment Conditions \\
\hline 
Normal$(\mu,\sigma^2)$  & $u$       & $\sigma^2$    & Covariates have finite second moments
\\
Poisson$(\lambda)$      & $e^u$     & $m$           & $X$ has a finite MGF in a neighbourhood of $2 \beta_0$
\\
Poisson$(\lambda)$      & $u^2$     & $m$           & Covariates have finite fourth moments
\\
Bernoulli$(\pi)$        & $F(u)$    & $m(1-m)$      & Covariates have finite fourth moments
\\
Gamma($\mu,k$)          & $e^u$     & $m^2/k$       & \begin{tabular}{@{}l@{}} $X$ has a finite MGF in a neighbourhood of $4 \beta_0$, \textbf{and} \\ the covariates have finite second moments \end{tabular}
\\
IG($\mu,\lambda$)       & $e^u$     & $m^3/\lambda$ & $X$ has a finite MGF in neighbourhoods of $6 \beta_0$ and $-\beta_0$
\\
NB($\mu, k$)            & $e^u$     & $m + m^2/k$   & \begin{tabular}{@{}l@{}} $X$ has a finite MGF in a neighbourhood of $4 \beta_0$, \textbf{and} \\ the covariates have finite third moments \end{tabular}
\\
\hline
\end{tabular}
\end{center}
\caption{Moment conditions on $X$ for the consistency of $\Sigma_n$. The conditions in this table are for when the dispersion parameter is known, i.e., for the Normal, gamma, inverse Gaussian, and negative binomial distributions when $\sigma^2$, $k$, $\lambda$, and $k$ are known.}
\label{tab:moment_conditions}
\end{table}

\bigskip

\noindent\textbf{Proof:}  The estimator $\Sigma_n$ can be written in the form
$$
A_n(\beta_n) - B_{n}(\beta_n) C_n^{-1}(\beta_n) B_{n}^\top(\beta_n),
$$
where $A_n(\beta)$ is a $G \times G$ matrix, $B_{n}(\beta)$ is a $G \times d$ matrix, and $C_n(\beta)$ is $d \times d$. These matrices are given by
\begin{align*}
A_n(\beta) &= \frac{1}{n} G_n(\beta) V_n(\beta) {G_n(\beta)}^\top,
\\
B_{n}(\beta) & = \frac{1}{n} G_n(\beta)V_n^{1/2}(\beta)W_n^{1/2}(\beta)X^*, \text{ and}
\\
C_n(\beta) & = \frac{1}{n} X^{*\top} W_n(\beta) X^*.
\end{align*}
The matrices $G_n(\beta)$, $V_n(\beta)$, and $W_n(\beta)$ are the same as $G_n^*$, $V_n^*$, and $W_n$ defined in the main text, except that the latter matrices are evaluated at $\beta_n$. Here we have emphasized the dependence of each entry on $\beta$; we will show that under our conditions each of $A_n$, $B_n$, and $C_n$ converges to its expected value uniformly in $\beta\in\overline{{\cal N}}$ where $\overline{\cal N} = \{\beta:\|\beta-\beta_0\|\le \delta \}$ with $\delta$ from Condition (E).  We will also show those limits are continuous functions of $\beta$. This will finish our proof of consistency.  

Each of these three matrices can be written in the form
$$
\frac{1}{n} \sum_{i=1}^n H(X_i,\beta).
$$
Under our assumptions, the matrix valued functions $H$ involved have finite expectations for all $\beta\in\overline{{\cal N}}$.  Our proof then uses Glivenko-Cantelli theorems, that is, uniform laws of large numbers; see the  Lemmas \ref{lemma:GC} and \ref{lemma:Donsker} below.

\bigskip

\noindent\textbf{Consistency of $A_n$}

\medskip

For the matrix $A_n$ the  $g,g'$ entry in $H(x,\beta)$
is
$$
1(k_{g-1} < \beta^\top x \le k_g)v(m(\beta^\top x))1(k_{g'-1} < \beta^\top x \le k_{g'}),
$$
which vanishes unless $g=g'$, in which case it is simply
$$
1(k_{g-1} < \beta^\top x \le k_g)v(m(\beta^\top x)).
$$
We apply Lemma~\ref{lemma:Donsker}. For $u\in\reals$ and $\beta\in \reals^d$ define the function $f_{\beta,u}$ by
$$
f_{\beta,u}(x)= v(m(\beta^\top x)) 1(\beta^\top x \le u),
$$
and the class of functions ${\cal F}_A$ by
$$
{\cal F}_A = \{f_{\beta,u}:  \|\beta-\beta_0\|\le \delta, u\in\reals\}.
$$
We apply the Lemma with $\beta^*=\beta_0$, $h(\beta,x) = v(m(\beta^\top x))$, and $M=M_A$. 

With these choices, conditions i-iii of the Lemma come immediately from Condition (E). We conclude that ${\cal F}_A$ is $P$-Glivenko-Cantelli. That is,
$$
\sup\left\{\left|\frac{1}{n} \sum_{i=1}^n f_{\beta,u}(X_i)- \Exp(v(m(\beta^\top X))1(\beta^\top X \le u))\right|;\beta\in \overline{\cal N}, u\in\reals\right\} \to 0
$$
almost surely. Let 
$$
J(\beta,u) = \Exp\left\{v(m(\beta^\top X)) 1(\beta^\top X \le u)\right\},
$$
and
$$
J_n(\beta,u) =\frac{1}{n}
\sum_{i=1}^n v(m(\beta^\top X_i)) 1(\beta^\top X_i \le u).
$$
The $g,g$ entry in $A_n$ is 
$$
J_n(\beta_n,k_g) - J_n(\beta_n,k_{g-1}).
$$
We have shown that
$$
\left\{J_n(\beta_n,k_g) - J_n(\beta_n,k_{g-1})\right\}-
\left\{J(\beta_n,k_g) - J(\beta_n,k_{g-1})\right\} \to 0
$$
almost surely.
Consistency of $A_n$ then follows from continuity in $\beta$ of $J(\beta,k_g)$ (for all $g$).
The Dominated Convergence Theorem shows that for any deterministic sequence $\beta_n$ converging to $\beta_0$ we have
$$
\lim_{n\to\infty} 
\Exp\left\{h(\beta_n,X)1(\beta_n^\top X \le k_g)\right\} =  
\Exp\left\{h(\beta_0,X)1(\beta_0^\top X \le k_g)\right\},
$$
provided $P(\beta_0^\top X = k_g) = 0$. This last follows from Condition (C).

\bigskip

\noindent\textbf{Consistency of $B_n$}

\medskip

For the matrix $B_n$ the  $g,j$ entry in $H(x_i,\beta)$ 
is
$$
1(k_{g-1} < \beta^\top x_i \le k_g) m'(\beta^\top x_i) x_{ij}.
$$

As for $A_n$ we define, for $u\in\reals$, $\beta\in \reals^d$, and $j\in \{1,\ldots,d\}$, the function $f_{\beta,u,j}$ to be the $j$th component of 
$$
f_{\beta,u}(x) = m'(\beta^\top x) 1(\beta^\top x \le u) x.
$$
The argument for 
$A_n$ together with the assumption on $M_B$ may be followed to prove that $B_n(\beta_n)$ converges almost surely to its expectation.
\bigskip

\noindent\textbf{Consistency of $C_n$}

\medskip

Finally we consider the matrix $C_n$  and show that $C_n(\beta_n)$ is a consistent estimator of the Fisher information matrix for a single observation.  
This matrix has the form
$$
C_n(\beta) = \frac{1}{n}\sum_{i=1}^n  X_i X_i^\top\frac{(m'(\beta^\top X_i))^2}{v(m(\beta^\top X_i))}.
$$
Define
$$
J_C(\beta) = \Exp\left(XX^\top\frac{(m'(\beta^\top X))^2}{v(m(\beta^\top X))}\right),
$$
and observe that $J_C(\beta_0) =I_1(\beta_0)$, the Fisher Information matrix. (It is a consequence of this observation and Condition (E) that $J_C(\beta)$ is finite for each $\beta\in \overline{{\cal N}}$.) We now apply Lemma~\ref{lemma:GC}; the Lemma is applied to the $j,j'$ component of $C_n$ but we can use the same bounding function for all components. In the Lemma we take $h(\beta,x)$ to be the $j,j'$ entry in
$$
 x x^\top \frac{(m'(\beta^\top x))^2}{v(m(\beta^\top x))}
$$
and  $M(x) =M_{C,\delta}(x)$. \hfill$\bullet$

\bigskip

\noindent\textbf{Verifying Condition (E)}

\medskip
It remains to interpret Condition (E) for each combination of inverse link $m$ and each variance function $v$ considered in this paper. We run through our models, which determine $v$, and then consider those $m$ which might be used for that model. We first present a table of link and variance combinations, along with the functions appearing in our conditions.
\begin{table}[ht]
\begin{center}
\begin{tabular}{lcccccc}
\hline
Distribution            & $m(u)$    & $v(m)$        & $m'(u)$   & $m'(u)v'(m(u))$    & $m''(u)$    & $(m'(u))^2/v(m(u))$ \\
\hline 
Normal$(\mu,\sigma^2)$  & $u$       & $\sigma^2$    & 1         & 0                 & 0             & $1/\sigma^2$
\\
Poisson$(\lambda)$      & $e^u$     & $m$           & $e^u$     & $e^u$             & $e^u$         & $e^u$
\\
Poisson$(\lambda)$      & $u^2$     & $m$           & $2u$      & $2u$              & $2$           & $4$
\\
Bernoulli$(\pi)$        & $F(u)$    & $m(1-m)$      & $f(u)$    & $(1-2F(u))f(u)$   & $f'(u)$       & $f^2(u)/[F(u)(1-F(u))]$
\\
Gamma($\mu,k$)          & $e^u$     & $m^2/k$       & $e^u$     & $2e^{2u}/k$       & $e^u$         & $k$
\\
IG($\mu, \lambda$)      & $e^u$     & $m^3/\lambda$ & $e^u$    & $3 e^{3u}/\lambda$  & $e^u$        & $\lambda/e^u$
\\
NB($\mu, k$)            & $e^u$     & $m+m^2/k$     & $e^u$    & $e^u + 2 e^{2u}/k$  & $e^u$          & $ e^u/(1+e^u/k)$
\\
\hline
\end{tabular}
\end{center}
\caption{Several distribution and link function combinations}
\end{table}

\bigskip

\noindent\textbf{Normal Family}:  Here $v(m) \equiv \sigma^2$ and we may take $M_{A,\delta}(x) =M= 0$ for any link and any $\delta>0$.  The most common link for the normal family is the identity for which $m'\equiv 1$, $m''\equiv 0$, and $M_{B,\delta}=0$. In this case Condition (E) reduces to assuming that the covariates have finite second moments. The conditions on $v(m(\beta_0^\top X))$ and on $m'(\beta_0^\top X)$ are both trivial.

\bigskip

\noindent\textbf{Poisson Family}: Here $v(m) = m$. Common links are the log link ($m(u) = e^u$) and the square root link ($m(u) = u^2$).

For the log link we find that 
$v(m(u)) =e^{u}$ so that
$$
M_{A,\delta}(x) = \|x\| \sup\{e^{\beta^\top x}: \|\beta-\beta_0\|\le \delta \},
$$
which means our condition is that $X$ has a finite moment generating function in a neighbourhood of $2 \beta_0$. We also have 
$$
m''(u) = m'(u) = e^u,
$$
and
$$
M_{B, \delta}(x) = \|x\|^2 \sup\{e^{\beta^\top x}: \|\beta-\beta_0\|\le \delta \}.
$$
The condition is the same as the one associated with $M_{A,\delta}$.
Finally,
$$
\frac{d}{du} \frac{(m'(u))^2}{v(m(u))}=e^u,
$$
so our condition is that $X$ has a finite moment generating function in a neighbourhood of $\beta_0$. We also need $\Exp(e^{2 \beta_0^\top X}) < \infty$ and $\Exp(e^{2 \beta_0^\top X} \|X\|^2) < \infty$. These lead to the same condition as the one associated with both $M_{A, \delta}$ and $M_{B, \delta}$. That is: for the Poisson family with log link, Condition (E) holds as long as $X$ has a finite moment generating function in a neighbourhood of $2\beta_0$.

For the square root link we have $m(u) = v(m(u))= u^2$. We find that
$$
w_A(u) = 2u.
$$
Thus,
$$
M_{A,\delta}(x) =2\|x\| \sup\{|\beta^\top x|: \| \beta-\beta_0 \| \le \delta\}.
$$
Writing $\beta$ in the form $\beta_0+a x + b v$ with $v$ any vector perpendicular to $x$ shows that we can regard $b=0$, and then 
$$
M_{A,\delta}(x) =2\|x\| \sup\{|\beta_0^\top x| +a \|x\|^2: |a|\|x\| \le \delta\}.
$$
The maximum of this piecewise linear function of $a$ that occurs in the braces must occur on the boundary of the interval imposed on $a$, so it is easily checked that
$$
M_{A,\delta}(x) =2\|x\|\left(|\beta_0^\top x | + \delta \|x\|\right).
$$
Thus, $M_{A,\delta}$ is square integrable if $X$ has 4 finite moments. We see that
$$
m''(u) = 2,
$$
and thus
$$
M_{B,\delta}(x) = 2\|x\|^2,
$$
which requires 4 finite moments. Finally, it is easily checked that for this link
$$
C_n(\beta) =\frac{4}{n} \sum_{i=1}^n X_i X_i^\top,
$$
which does not depend on $\beta$. For the Poisson family with a square root link, we also require $\Exp((\beta_0^\top X)^4) < \infty$ and $\Exp(4 (\beta_0^\top X)^2 \|X\|^2) < \infty$, but these do not add any further moment conditions. That is: for the Poisson family with square root link, Condition (E) holds as long as $X$ has 4 finite moments.

\bigskip

\noindent\textbf{Bernoulli Family}: For the Bernoulli($\theta$) model we have  $v(m) = m(1-m)$. The inverse links we consider all have the form
$$
m(u) = F(u),
$$
for a smooth cdf $F$ with corresponding smooth density $f$.  We therefore have
$$
v(m(u)) = F(u)(1-F(u)),
$$
which is bounded by $1$, and
$$
m'(u) = f(u).
$$
For all 4 links (logit, probit, cauchit, and cloglog) we find that there is a constant, $C$, such that, for all $u$,
$$
f(u)\le C.
$$
For all 4 links there is also another constant, $C_2$, such that, for all $u$,
$$
|f'(u)| \leq C_2.
$$
Therefore,
$$
|w_A(u)| = |f(u) (1-2F(u))| \leq C,
$$
and so
$$
M_{A,\delta}(x) \leq \|x\| C,
$$
which amounts to 2 finite moments. Also,
$$
M_{B, \delta}(x) \leq C_2 \|x\|^2,
$$
which amounts to 4 finite moments. 

The quantity
$$
|w_C(u)| = \left|\frac{d}{du} \frac{\left(m'(u)\right)^2}{v(m(u))}\right|= \left|\frac{2 f(u)f'(u)}{F(u)(1-F(u))}-
\frac{f^3(u)(1-2F(u))}{\left[F(u)\left\{1-F(u)\right\}\right]^2}\right|
$$
is bounded by some constant. Therefore,  our condition on $M_{C, \delta}(x)$ amounts to requiring three finite moments for $X$.

\bigskip

\noindent\textbf{Gamma Family}: Here $v(m) = m^2/k$, where $k>0$ is the shape parameter in the Gamma distribution. We consider the log link, $m(u) = e^u$. The square root link, $m(u) = u^2$, is included to highlight how to deal with other link functions.  

For the log link we find that 
$v(m(u)) =e^{2u}/k$, and $m'(u) = e^u$, so that
$$
m'(u) v'(m(u))  = 2e^{2u}/k.
$$
Thus,
$$
M_{A,\delta}(x) = 2 \|x\|\sup\left\{e^{2\beta^\top x}: \|\beta-\beta_0\|\le \delta\right\}/k,
$$
which means our condition on $M_{A,\delta}$ is that $X$ has a finite moment generating function in a neighbourhood of $4\beta_0$. We also take 
$$
M_{B,\delta}(x) = \|x\|^2\sup\left\{e^{\beta^\top x}: \|\beta-\beta_0\|\le \delta\right\},
$$
which leads to the strictly weaker condition that $X$ has a finite moment generating function in a neighbourhood of $2\beta_0$. 
Finally,
$$
\frac{(m'(u))^2}{v(m(u))} = k,
$$
whose derivative is 0 so we need only 
$$
M_{C,\delta}(u) = 0.
$$
Thus, $C_n(\beta)$ does not depend on $\beta$, so consistency is a consequence of two finite moments for $X$. The conditions on $v(m(\beta_0^\top X))$ and on $m'(\beta_0^\top X)$ are the same as those above; they amount to a finite moment generating function of $X$ at $4\beta_0$  and at $2\beta_0$.

For the square root link, we have $v(m)=m^2/k$, $m(u)=u^2$, and $v(m(u))= u^4/k$. We find
$$
m'(u)v'(m(u))=4u^3/k.
$$
Thus,
$$
M_{A,\delta}(x) = 4\|x\|\left(|\beta_0^\top x| + \delta \|x\|\right)^3/k,
$$
which means we require $X$ to have 8 finite moments. Evidently, $ m''(u) = 2$. Thus, we may take
$$
M_{B,\delta}(x) = 2 \|x\|^2,
$$
which leads to a weaker condition, namely, 4 finite moments for $X$. Finally,
$$
\frac{(m'(u))^2}{v(m(u))}= \frac{4u^2 k}{u^4} = 
\frac{4 k}{u^2}.
$$
This function diverges at $u=0$, so we need to add an assumption: there is an $\epsilon>0$ and a $\delta>0$ such that, for all $\beta$ with $\|\beta-\beta_0\|\le \delta$, we have
$$
P( \beta^\top X \ge \epsilon ) = 1 .
$$
In this case, 
$$
\frac{d}{du} \frac{(m'(u))^2}{v(m(u))} = \frac{-8 k}{u^3},
$$
and $M_{C,\delta}(x) \leq 8 \|x\|^3 k/\epsilon^3$. This amounts to three finite moments for $X$. The conditions on $v(m(\beta_0^\top X))$ and on $m'(\beta_0^\top X)$ amount to  $\Exp((\beta_0^\top X)^8) < \infty$ and $\Exp((\beta_0^\top X)^2 \|X\|^2) < \infty$, i.e., 8 finite moments which matches the requirement for $M_{A, \delta}$.

\bigskip

\noindent\textbf{Inverse Gaussian Family}: Here $v(m) = m^3/\lambda$. For the log link, $m(u) = e^u$, we find that 
$v(m(u)) = e^{3u}/\lambda$ so that
$$
M_{A,\delta}(x) = 3 \|x\| \sup\{e^{3\beta^\top x}: \|\beta-\beta_0\|\le \delta\}/\lambda,
$$
which means our condition on $M_{A,\delta}$ is that $X$ has a finite moment generating function in a neighbourhood of $6\beta_0$. We also have 
$$
m''(u)=m'(u) =e^u,
$$
and therefore take 
$$
M_{B,\delta}(x) = \|x\|^2 \sup\{e^{\beta^\top x}: \|\beta-\beta_0\|\le \delta\},
$$
which leads to the strictly weaker condition that $X$ has a finite moment generating function in a neighbourhood of $2\beta_0$. 
Finally,
$$
\frac{(m'(u))^2}{v(m(u))}= \frac{\lambda}{e^u},
$$
and so
$$
\frac{d}{du} \frac{(m'(u))^2}{v(m(u))} = \frac{d}{du} \frac{\lambda}{e^u} = \frac{-\lambda}{e^u}.
$$
Thus,
$$
M_{C,\delta}(u) = \lambda \|x\|^3 \cdot \sup\{e^{-\beta^\top x}: \|\beta-\beta_0\|\le \delta\}.
$$
Our condition is that $X$ has a finite moment generating function in some neighbourhood of $-\beta_0$. We also need $\Exp(e^{6 \beta_0^\top X}) < \infty$, and $\Exp(\|X\|^2 e^{2 \beta_0^\top X}) < \infty$. Our overall condition is therefore that $X$ has a finite moment generating function in some neighbourhood of $6 \beta_0$ and in some neighbourhood of $-\beta_0$. We remark that the set of $\beta$ where a moment generating function is finite is convex and will include both these neighbourhoods and a tube containing them.

\bigskip

\noindent\textbf{Negative Binomial Family}: We have $v(m)=m+m^2/k$, with $k>0$. We consider the log link, although the square root and identity links are also sometimes used. For the log link, $m(u) = e^u$, and 
$$
|w_A(u)| = e^u + \frac{2e^{2u}}{k},
$$
so that
$$
M_{A, \delta}(x) = \|x\| \sup\{e^{\beta^\top x} + 2e^{2\beta^\top x}/k, \|\beta-\beta_0\| \le \delta\}.
$$
Our condition then amounts to $X$ having a finite moment generating function in a neighbourhood of $4\beta_0$. Also,
$$
M_{B, \delta}(x) = \|x\|^2 \sup\{e^{\beta^\top x}, \|\beta-\beta_0\|\le \delta\},
$$
which leads to the strictly weaker condition of $X$ having a finite moment generating function in a neighbourhood of $2\beta_0$. For $C_n$, we see that
$$
\left|\frac{d}{du} \frac{\left(m'(u)\right)^2}{v(m(u))}\right| = \frac{e^u}{(1+e^u/k)^2} \leq C,
$$
for some constant $C > 0$. Therefore, our condition associated with $M_{C, \delta}(x)$ is that $X$ has three finite moments. We also require $\Exp(v^2(m(\beta_0^\top X))) < \infty$, $\Exp(e^{2 \beta_0^\top X} \|X\|^2) < \infty$, and $\Exp(\|X\|^2)<\infty$, but these hold, provided that the above conditions on the moment generating function of $X$ are satisfied.

\bigskip

The proofs above used two lemmas, which in turn depend on a third. All of the lemmas synthesize results in \cite{kosorok2007introduction}.

\bigskip

\begin{lemma} \label{lemma:GC} Suppose $X_1,X_2,\ldots$ are iid $d$-dimensional vectors with the same distribution as $X$, and $\beta$ is a $d$-dimensional vector taking values in some open set $O$. Suppose $h$ is a real valued function of the pair $(x,\beta)$ which is continuously differentiable in $\beta$ for each $x$.
Let $K$ be some compact subset of $O$ with diameter denoted by diam$(K)$.  Assume:
\begin{enumerate}
    \item there is some $\beta^*\in K$ such that 
$$
\Exp\left\{|h(X,\beta^*)|\right\} < \infty,
$$

\item there is a function $M(x)$ such that for all $x$ in the support of $X$ and all $\beta\in K$
    $$
    \left\|\frac{\partial}{\partial \beta}h(x,\beta) \right\| \le M(x),
    $$ and,

\item the random variable $M(X)$ is  integrable:
    $$
    \Exp(M(X)) < \infty.
    $$
\end{enumerate}
Then,
\begin{enumerate}

\item The family of functions
    $$
    {\cal F}_{h,K} = \left\{x\mapsto h(\beta,x); \beta\in K \right\}
    $$
    is pointwise measurable.  
    
    \item This family has bounded uniform entropy integral with respect to the square integrable envelope 
    $$
    M^*(x) = \sqrt{2}\sqrt{M^2(x) + \left\{ {\rm diam}(K) h(\beta^*,x)\right\}^2} .
    $$
       
    \item The class
    ${\cal F}_{h,K}$ is $P$-Glivenko-Cantelli, that is, 
$$
\sup_{\beta\in K, u \in \reals^d} \left|\frac{1}{n} \sum_{i=1}^n h(X_i,\beta) - \Exp\left\{h(X,\beta)\right\}\right| \to 0
$$
almost surely. 

\item The map
$$
\beta \mapsto \Exp\left\{h(X,\beta)\right\}
$$
is uniformly continuous on $K$.

\end{enumerate}

\end{lemma}

\noindent\textbf{Proof}:  

\medskip

To prove the first statement we must find a countable subset ${\cal G}$ of ${\cal F}_{h,K}$ such that every $f \in {\cal F}_{h,K} $ is the pointwise limit of a sequence of elements of ${\cal G}$. We find ${\cal G}$ by a route which helps with our proof below of the second and third statements.

A set of closed balls of radius $\delta$ covers  a set $B \subset O$ if $B$ is contained in the union.  The covering number of $B$, denoted by 
$N(\delta,B,\|\cdot\|)$ is the smallest integer $N$ for which there is a set $\beta_1,\ldots,\beta_N$ of elements of $B$ such that for every $\beta\in B$ there is a $j$ with $\|\beta-\beta_j\| \le \delta$. The set of such balls is said to $\delta$-cover $B$.  We now bound, by a standard volume argument,
the value of $N$ when $B$ is a ball.

The volume of a ball of radius $r$ in $\reals^d$ is proportional to $r^d$.  Thus if there are $N$ disjoint balls of radius $\epsilon$ in $\reals^d$ all of which lie in some ball of radius $R$ then the volume of those $N$ small balls is $N$ times the volume of a single one and less than the volume of the ball of radius $R$. So $N \le (R/\epsilon)^d$. Consider now a set of such balls of maximal size; this collection is said to pack $B$ and the corresponding value of $N$ is the $\epsilon$ packing number. The collection of $N$ balls with the same centers but radius $2\epsilon$ contains the ball of radius $R$ for if not we could fit in another ball of radius $\epsilon$. 
So for any ball $B$ of radius $R$ we get
$$
N(\epsilon,B,\|\cdot\|) \le (2R/\epsilon)^d.
$$

For each $\epsilon>0$ we have identified a finite set say $B_\epsilon$ of points in $K$ such that every point in $K$ is within $\epsilon$ of some member $\beta$ of $B_\epsilon$.  Take $B$ to be the union over positive integers $n$ of $B_{1/n}$ and let ${\cal G}$ be the corresponding elements of ${\cal F}_{j,K}$.  Evidently $B$ and ${\cal G}$ are countable and $B$ is dense in  $K$.  Every $\beta\in K$ is thus the limit of a sequence of points $\beta_n\in B$ and the corresponding $f_\beta$ is the pointwise limit of $f_{\beta_n}$ because $h$ is continuous in $\beta$.  This proves Statement 1.

Now we turn to the second  statement.
Take $B$ to be a ball of radius $R \le {\rm diam}(K)$ which contains $K$.  Find $\beta_1,\ldots,\beta_N$ in $B$ so that $N=N(\epsilon,B,\|\cdot\|)$ and the $N$ balls centered at the $\beta_j$ having radius $\epsilon$ cover $B$. For each such ball which intersects $K$ let $\beta_j^*$ be in the intersection. Every point in $K$ is within the union of the balls centered at those $\beta_j$ for which the intersection with $K$ is not empty. So every point in $K$ is within balls centred at $\beta_j^*$ but with radius $2\epsilon$. Thus, 
$$
N(\epsilon,K,\|\cdot\|) \le (4{\rm diam}(K)/\epsilon)^2.
$$

Now suppose that $\beta\in K$. Fix $\epsilon>0$ and find $N=N(\epsilon,K,\|\cdot\|)$ points, say $\{\beta_1,\ldots,\beta_N\}$, in $K$ such that $K$ is contained in union of the $N$ balls of radius $\epsilon$ centered at the $\beta_j$.  
For each $f\in {\cal F}_{h,K}$ we have $f=f_\beta$ for some $\beta\in K$. Find $j$ so that $\| \beta-\beta_j \|\le \epsilon$. If $Q$ is a finite discrete measure on the support of $X$ there is a set of points $x_1,\ldots,x_k$ and corresponding probabilities $q_1,\ldots,q_k$ such that $\sum_1^k q_i = 1$ and $Q(X=x_i)=q_i$ for $j=1,\ldots,k$. Now
$$
\left| f_\beta(x_i)-f_{\beta_j}(x_i) \right| \le \|\beta-\beta_j\| M(x_i),
$$
by Taylor's theorem. Thus, the $L_1(Q)$ norm of $f_\beta-f_{\beta_j}$ satisfies
\begin{align*}
\|f_\beta-f_{\beta_j}\|_{Q,1} &=
\sum_{i=1}^k |f_\beta(x_i) -f_{\beta_j}(x_i)| q_i
\\
& \le \|\beta-\beta_j\| \sum_{i=1}^k |M(x_i)| q_i 
\\
& \le \epsilon \sum_{i=1}^k |M^*(x_i)| q_i 
\\
& \le \epsilon \|M^*\|_{Q,1}.
\end{align*}
Increasing the first argument in the covering number cannot increase the covering number itself.
Thus,
$$
N\left(\epsilon\|M^*\|_{Q,1},{\cal F},L_1(Q)\right) \le N\left(\epsilon,K,\|\cdot\|\right)
\le 
(4{\rm diam}(K)/\epsilon)^2 < \infty.
$$
This proves the bound on the uniform covering numbers
\begin{equation}\label{eq:uniformentropy}
\sup_{Q}N(\epsilon\|M\|_{Q,1},{\cal F},L_1(Q)) \le N(\epsilon,K,\|\cdot\|)
\le (4{\rm diam}(K)/\epsilon)^2 < \infty.
\end{equation}
This is Statement 2.

Since pointwise measurability implies $P$-measurability for every $P$, we have verified all the conditions of Theorem 8.14, page 145 in
 \cite{kosorok2007introduction}. Statement 3 follows.

The fourth assertion of the lemma is an application of the Dominated convergence theorem. If the sequence $\beta_n$ converges to some $\beta\in K$ then 
$|f_{\beta_n}(x)-f_\beta(x)| <  \sup_n \|\beta_n-\beta\|M(x)$ and the right hand side of this inequality is integrable. Uniform continuity is automatic because $K$ is compact. (Indeed, the assumptions on $M$ guarantee that this expectation is a differentiable function of $\beta$ on the interior of $K$.)\hfill $\bullet$

Our second Lemma deals with processes involving indicators. It deduces Glivenko-Cantelli results from Donsker results; it seems likely that this contributes to an increase in the strength of our moment conditions.

\begin{lemma} \label{lemma:Donsker} Suppose $X_1,X_2,\ldots$ are iid $d$-dimensional vectors with the same distribution as $X$, and $\beta$ is a $d$-dimensional vector taking values in some open set $O$. Suppose $h$ is a real valued function of the pair $(x,\beta)$ which is continuously differentiable in $\beta$ for each $x$.
Let $K$ be some compact subset of $O$ with diameter denoted by diam$(K)$.  Assume:
\begin{enumerate}[i)]
    \item there is some $\beta^*\in O$ such that 
$$
\Exp\left\{h^2(X,\beta^*)\right\} < \infty,
$$

\item there is a function $M(x)$ such that for all $x$ in the support of $X$ and all $\beta\in K$
    $$
    \left\|\frac{\partial}{\partial \beta}h(x,\beta) \right\| \le M(x),
    $$ and,

\item the random variable $M(X)$ is square integrable:
    $$
    \Exp\left(M^2(X)\right) < \infty.
    $$
\end{enumerate}
Then, 
\begin{enumerate}
 
\item The class of functions
 $$
    {\cal F}_{h,K} = \left\{f_\beta: f_\beta(x) = h(\beta,x), \beta\in K\right\}
 $$ 
    has square integrable envelope
    $$
    M^*(x) = \sqrt{2} \sqrt{\left\{{\rm diam}(K) M(x)\right\}^2+h^2(\beta^*,x)}.
    $$
    
\item The class of functions
    ${\cal F}_{h,K}$
    is pointwise measurable.

\item The family of functions
    \begin{equation}\label{eq:definecalFI}
    {\cal F}_I \equiv \{x\mapsto 1(\beta^\top x \le u), \beta\in \reals^d, u\in\reals\}
    \end{equation}
    is $P$-measurable for any $P$ and has Vapnik-Chervonenkis dimension $d+2$. 
    This family has bounded uniform entropy integral with envelope 1. 
    
\item The family
$$
{\cal F}_A = \left\{f: f= f_1 f_2, f_1 \in {\cal F}_I, f_2 \in {\cal F}_{h,K}\right\}
$$
has bounded uniform entropy integral with respect to the envelope $M^*$. This envelope is square integrable.

\item The family ${\cal F}_A$ is $P$-measurable for any $P$. For each $0<\delta\le\infty$ the family
$$
{\cal F}_{A,\delta}= \left\{f-g:f,g\in{\cal F}_A, \Exp\left\{f(X)-g(X)\right\}^2<\delta^2\right\}
    $$  
is $P$-measurable for any $P$. The family 
$$
{\cal F}_{A,\infty}^2=  \left\{(f-g)^2:f,g\in{\cal F}_A\right\}
$$
is $P$ measurable for all $P$.

\item The family ${\cal F}_A$ is $P$-Donsker.

\item The family ${\cal F}_{h,u,K}$ is $P$-Donsker.

\item The family ${\cal F}_{h,u,K}$ is $P$-Glivenko-Cantelli.

\end{enumerate}

\end{lemma}

\noindent\textbf{Proof}:  

\medskip

The first statement is elementary.
The second statement is contained in Lemma~\ref{lemma:GC}.
The third statement is Lemma 9.12 on page 161 of \cite{kosorok2007introduction}.
 
The fourth statement is a consequence of Theorem 9.15 of \cite{kosorok2007introduction} which asserts that given two classes ${\cal F}_1$ and ${\cal F}_2$ which have bounded entropy integral with envelopes $M_1$ and $M_2$ the class of all products of a function from each has a bounded uniform entropy integral with envelope the product $M_1M_2$. 

The fifth statement follows from a small extension of Lemma 8.12 on page 143 in \cite{kosorok2007introduction}; see Lemma~\ref{lemma:Pmeasurable} below.

The sixth statement is a consequence of the fourth and fifth and Theorem~8.19 on page 149 in \cite{kosorok2007introduction}.
Any subfamily of a $P$-Donsker family is $P$-Donsker; the seventh assertion follows. The final statement is the assertion that $P$-Donsker implies $P$-Glivenko-Cantelli which is contained in Lemma~8.17 on page 148.

\begin{lemma}\label{lemma:Pmeasurable}
    Let ${\cal G}$ be a pointwise measurable family of functions on $\reals^d$ and let ${\cal F}_I$ be the family in~(\ref{eq:definecalFI}). Then, the family 
    $$
    {\cal H} \equiv {\cal G}{\cal F}_I\equiv\left\{fg: f\in {\cal F}_I, g\in {\cal G}\right\}
    $$
    is $P$-measurable for all $P$.  Moreover, if we define for $0 < \delta< \infty$ the class
    $$
    {\cal H}_\delta = \left\{f-g:f,g\in{\cal H}, \Exp\left\{f(X)-g(X)\right\}^2<\delta^2\right\}
    $$  
    and the class
    $$
    {\cal H}_\infty^2 = \left\{(f-g)^2:f,g\in{\cal H}\right\},
    $$ 
    then all these classes are $P$-measurable for any $P$. 
    \end{lemma}
    
The proof of the Lemma is entirely analogous to that of Lemma 8.12 on page 143 in \cite{kosorok2007introduction}. \hfill $\bullet$

\setlength{\bibsep}{0pt plus 0.3ex} 
\bibliography{main.bib}

\begin{thebibliography}{38}
\providecommand{\natexlab}[1]{#1}
\providecommand{\url}[1]{\texttt{#1}}
\expandafter\ifx\csname urlstyle\endcsname\relax
  \providecommand{\doi}[1]{doi: #1}\else
  \providecommand{\doi}{doi: \begingroup \urlstyle{rm}\Url}\fi

\bibitem[Agresti(1996)]{agresti1996introduction}
Alan Agresti.
\newblock \emph{An introduction to categorical data analysis}.
\newblock John Wiley \& Sons, 1996.

\bibitem[Agresti(2015)]{agresti2015foundations}
Alan Agresti.
\newblock \emph{Foundations of linear and generalized linear models}.
\newblock John Wiley \& Sons, 2015.

\bibitem[Andrews(1987)]{andrews1987asymptotic}
Donald~W.K. Andrews.
\newblock Asymptotic results for generalized {W}ald tests.
\newblock \emph{Econometric Theory}, 3\penalty0 (3):\penalty0 348--358, 1987.

\bibitem[Becker et~al.(2021)Becker, Loughin, Astudillo-Sanchez, and
  Wethington]{becker2020hummingbird}
C.~D. Becker, T.~M. Loughin, E.~Astudillo-Sanchez, and S.~Wethington.
\newblock Interaction strengths between hummingbirds and flowers in {E}cuador.
\newblock \textit{The Wilson Journal of Ornithology}, Accepted, 2021.

\bibitem[Bilder and Loughin(2014)]{bilder2014analysis}
Christopher~R. Bilder and Thomas~M. Loughin.
\newblock \emph{Analysis of categorical data with R}.
\newblock Chapman and Hall/CRC, 2014.

\bibitem[Blizzard and Hosmer(2006)]{blizzard2006parameter}
Leigh Blizzard and David~W. Hosmer.
\newblock Parameter estimation and goodness-of-fit in log binomial regression.
\newblock \emph{Biometrical Journal}, 48\penalty0 (1):\penalty0 5--22, 2006.

\bibitem[Canary(2013)]{canary2013grouped}
Jana~D. Canary.
\newblock \emph{Grouped goodness-of-fit tests for binary regression models}.
\newblock PhD thesis, University of Tasmania, 2013.

\bibitem[Canary et~al.(2016)Canary, Blizzard, Barry, Hosmer, and
  Quinn]{canary2016summary}
Jana~D. Canary, Leigh Blizzard, Ronald~P. Barry, David~W. Hosmer, and
  Stephen~J. Quinn.
\newblock Summary goodness-of-fit statistics for binary generalized linear
  models with noncanonical link functions.
\newblock \emph{Biometrical Journal}, 58\penalty0 (3):\penalty0 674--690, 2016.

\bibitem[Cheng and Wu(1994)]{cheng1994testing}
K.F. Cheng and J.W. Wu.
\newblock Testing goodness of fit for a parametric family of link functions.
\newblock \emph{Journal of the American Statistical Association}, 89\penalty0
  (426):\penalty0 657--664, 1994.

\bibitem[Christensen and Lin(2015)]{christensen2015lack}
Ronald Christensen and Yong Lin.
\newblock Lack-of-fit tests based on partial sums of residuals.
\newblock \emph{Communications in Statistics - Theory and Methods}, 44\penalty0
  (13):\penalty0 2862--2880, 2015.

\bibitem[Davidov et~al.(2018)Davidov, Jelsema, and Peddada]{davidov2018testing}
Ori Davidov, Casey~M. Jelsema, and Shyamal Peddada.
\newblock Testing for inequality constraints in singular models by trimming or
  winsorizing the variance matrix.
\newblock \emph{Journal of the American Statistical Association}, 113\penalty0
  (522):\penalty0 906--918, 2018.

\bibitem[DeHart et~al.(2008)DeHart, Tennen, Armeli, Todd, and
  Affleck]{dehart2008drinking}
Tracy DeHart, Howard Tennen, Stephen Armeli, Michael Todd, and Glenn Affleck.
\newblock Drinking to regulate negative romantic relationship interactions: The
  moderating role of self-esteem.
\newblock \emph{Journal of Experimental Social Psychology}, 44\penalty0
  (3):\penalty0 527--538, 2008.

\bibitem[Fagerland and Hosmer(2013)]{fagerland2013goodness}
Morten~W. Fagerland and David~W. Hosmer.
\newblock A goodness-of-fit test for the proportional odds regression model.
\newblock \emph{Statistics in Medicine}, 32\penalty0 (13):\penalty0 2235--2249,
  2013.

\bibitem[Fagerland and Hosmer(2016)]{fagerland2016tests}
Morten~W. Fagerland and David~W. Hosmer.
\newblock Tests for goodness of fit in ordinal logistic regression models.
\newblock \emph{Journal of Statistical Computation and Simulation}, 86\penalty0
  (17):\penalty0 3398--3418, 2016.

\bibitem[Fagerland et~al.(2008)Fagerland, Hosmer, and
  Bofin]{fagerland2008multinomial}
Morten~W. Fagerland, David~W. Hosmer, and Anna~M. Bofin.
\newblock Multinomial goodness-of-fit tests for logistic regression models.
\newblock \emph{Statistics in Medicine}, 27\penalty0 (21):\penalty0 4238--4253,
  2008.

\bibitem[Fahrmeir and Kaufmann(1985)]{fahrmeir1985consistency}
Ludwig Fahrmeir and Heinz Kaufmann.
\newblock Consistency and asymptotic normality of the maximum likelihood
  estimator in generalized linear models.
\newblock \emph{The Annals of Statistics}, pages 342--368, 1985.

\bibitem[Fahrmeir and Kaufmann(1986)]{fahrmeir1986correction}
Ludwig Fahrmeir and Heinz Kaufmann.
\newblock Correction: consistency and asymptotic normality of the maximum
  likelihood estimator in generalized linear models.
\newblock \emph{The Annals of Statistics}, 14\penalty0 (4):\penalty0
  1643--1643, 1986.

\bibitem[Gonz{\'a}lez-Manteiga and Crujeiras(2013)]{gonzalez2013updated}
Wenceslao Gonz{\'a}lez-Manteiga and Rosa~M. Crujeiras.
\newblock An updated review of goodness-of-fit tests for regression models.
\newblock \emph{Test}, 22\penalty0 (3):\penalty0 361--411, 2013.

\bibitem[Halteman(1980)]{halteman1980goodness}
William~A. Halteman.
\newblock A goodness of fit test for binary logistic regression.
\newblock \emph{Unpublished doctoral dissertation, Department of Biostatistics,
  University of Washington, Seattle, WA}, 1980.

\bibitem[Hosmer and Hjort(2002)]{hosmer2002goodness}
David~W. Hosmer and Nils~L. Hjort.
\newblock Goodness-of-fit processes for logistic regression: simulation
  results.
\newblock \emph{Statistics in Medicine}, 21\penalty0 (18):\penalty0 2723--2738,
  2002.

\bibitem[Hosmer and Lemeshow(1980)]{hosmer1980goodness}
David~W. Hosmer and Stanley Lemeshow.
\newblock Goodness of fit tests for the multiple logistic regression model.
\newblock \emph{Communications in Statistics - Theory and Methods}, 9\penalty0
  (10):\penalty0 1043--1069, 1980.

\bibitem[Klein and Moeschberger(1997)]{klein2006survival}
John~P Klein and Melvin~L Moeschberger.
\newblock \emph{Survival analysis: techniques for censored and truncated data}.
\newblock Springer Science \& Business Media, 1997.

\bibitem[Kosorok(2007)]{kosorok2007introduction}
M.R. Kosorok.
\newblock \emph{Introduction to Empirical Processes and Semiparametric
  Inference}.
\newblock Springer Series in Statistics. Springer New York, 2007.
\newblock ISBN 9780387749785.
\newblock URL \url{https://books.google.ca/books?id=FXaYjTQIUZ8C}.

\bibitem[Lemeshow and Hosmer(1982)]{lemeshow1982review}
Stanley Lemeshow and David~W. Hosmer.
\newblock A review of goodness of fit statistics for use in the development of
  logistic regression models.
\newblock \emph{American Journal of Epidemiology}, 115\penalty0 (1):\penalty0
  92--106, 1982.

\bibitem[Lemeshow et~al.(1988)Lemeshow, Teres, Avrunin, and
  Pastides]{lemeshow1988predicting}
Stanley Lemeshow, Daniel Teres, Jill~Spitz Avrunin, and Harris Pastides.
\newblock Predicting the outcome of intensive care unit patients.
\newblock \emph{Journal of the American Statistical Association}, 83\penalty0
  (402):\penalty0 348--356, 1988.

\bibitem[Lin et~al.(2002)Lin, Wei, and Ying]{lin2002model}
D.Y. Lin, L.J. Wei, and Z.~Ying.
\newblock Model-checking techniques based on cumulative residuals.
\newblock \emph{Biometrics}, 58\penalty0 (1):\penalty0 1--12, 2002.

\bibitem[Liu et~al.(2004)Liu, Meiring, and Wang]{liu2005testing}
Anna Liu, Wendy Meiring, and Yuedong Wang.
\newblock Testing generalized linear models using smoothing spline methods.
\newblock \emph{Statistica Sinica}, pages 235--256, 2004.

\bibitem[L{\"u}tkepohl and Burda(1997)]{lutkepohl1997modified}
Helmut L{\"u}tkepohl and Maike~M. Burda.
\newblock Modified {W}ald tests under nonregular conditions.
\newblock \emph{Journal of Econometrics}, 78\penalty0 (2):\penalty0 315--332,
  1997.

\bibitem[Moore and Spruill(1975)]{moore1975unified}
David~S. Moore and Marcus~C. Spruill.
\newblock Unified large-sample theory of general chi-squared statistics for
  tests of fit.
\newblock \emph{The Annals of Statistics}, pages 599--616, 1975.

\bibitem[Pulkstenis and Robinson(2002)]{pulkstenis2002two}
Erik Pulkstenis and Timothy~J. Robinson.
\newblock Two goodness-of-fit tests for logistic regression models with
  continuous covariates.
\newblock \emph{Statistics in Medicine}, 21\penalty0 (1):\penalty0 79--93,
  2002.

\bibitem[Quinn et~al.(2015)Quinn, Hosmer, and Blizzard]{quinn2015goodness}
Stephen~J. Quinn, David~W. Hosmer, and C.~Leigh Blizzard.
\newblock Goodness-of-fit statistics for log-link regression models.
\newblock \emph{Journal of Statistical Computation and Simulation}, 85\penalty0
  (12):\penalty0 2533--2545, 2015.

\bibitem[Rodr{\'\i}guez-Campos et~al.(1998)Rodr{\'\i}guez-Campos,
  Gonz{\'a}lez-Manteiga, and Cao]{rodriguez1998testing}
M.~Celia Rodr{\'\i}guez-Campos, Wenceslao Gonz{\'a}lez-Manteiga, and Ricardo
  Cao.
\newblock Testing the hypothesis of a generalized linear regression model using
  nonparametric regression estimation.
\newblock \emph{Journal of Statistical Planning and Inference}, 67\penalty0
  (1):\penalty0 99--122, 1998.

\bibitem[Stute and Zhu(2002)]{stute2002model}
Winfried Stute and Li-Xing Zhu.
\newblock Model checks for generalized linear models.
\newblock \emph{Scandinavian Journal of Statistics}, 29\penalty0 (3):\penalty0
  535--545, 2002.

\bibitem[Stute et~al.(1998)Stute, Thies, and Zhu]{stute1998model}
Winfried Stute, Silke Thies, and Li-Xing Zhu.
\newblock Model checks for regression: an innovation process approach.
\newblock \emph{The Annals of Statistics}, 26\penalty0 (5):\penalty0
  1916--1934, 1998.

\bibitem[Su and Wei(1991)]{su1991lack}
John~Q. Su and L.J. Wei.
\newblock A lack-of-fit test for the mean function in a generalized linear
  model.
\newblock \emph{Journal of the American Statistical Association}, 86\penalty0
  (414):\penalty0 420--426, 1991.

\bibitem[Surjanovic and Loughin(2021)]{surjanovic2021improving}
Nikola Surjanovic and Thomas~M. Loughin.
\newblock Improving the {H}osmer-{L}emeshow goodness-of-fit test in large
  models with replicated trials.
\newblock \emph{arXiv preprint arXiv:2102.12698}, 2021.

\bibitem[Tsiatis(1980)]{tsiatis1980note}
Anastasios~A. Tsiatis.
\newblock A note on a goodness-of-fit test for the logistic regression model.
\newblock \emph{Biometrika}, 67\penalty0 (1):\penalty0 250--251, 1980.

\bibitem[Xiang and Wahba(1995)]{xiang1995testing}
Dong Xiang and Grace Wahba.
\newblock Testing the generalized linear model null hypothesis versus
  ‘smooth’ alternatives.
\newblock Technical Report 953, Department of Statistics, University of
  Wisconsin, 1995.

\end{thebibliography}
\bibliographystyle{plainnat}

\end{document}